\documentclass[english,12pt,aps,prd,a4paper,preprintnumbers,floatfix,nofootinbib,showpacs,superscriptaddress, notitlepage]{revtex4-1}
\pdfoutput=1
\usepackage[english]{babel}
\usepackage{amsmath}
\usepackage{graphicx}
\usepackage{color}
\usepackage{mathrsfs}   
\usepackage{amssymb}
\usepackage{float}
\usepackage{enumerate}
\usepackage{caption}
\usepackage{subcaption}
\usepackage[colorlinks=true,citecolor=darkred,urlcolor=darkred, pdfborder={0 0 0}]{hyperref}
\definecolor{darkred}{rgb}{0.6,0,0}
\usepackage[normalem]{ulem} 
\usepackage{dcolumn}
\usepackage{bm}
\usepackage{graphicx}
\usepackage{multirow}
\usepackage{rotating}
\usepackage{subcaption}
\usepackage[sort&compress]{natbib}
\usepackage{epstopdf}
\usepackage[utf8]{inputenc} 

\newcommand {\be} {\begin{equation}}
\newcommand {\ee} {\end{equation}}
\newcommand {\hc} {{\rm h.c.}}

\definecolor{greenLinks}{rgb}{0, 0.6, 0} 
\definecolor{blueLinks}{rgb}{0, 0, 0.6}
\definecolor{redLinks}{rgb}{0.6, 0, 0}
\definecolor{tempText}{rgb}{0.55, 0.10,0.67}
\definecolor{eprintLinks}{rgb}{0.4, 0.4, 0.4}
\definecolor{journalLinks}{rgb}{0.6, 0, 0}

\def\21{$\mathrm{SU(2)_L \otimes U(1)_Y}$ }
\def\31{$\mathrm{SU(3)_c \otimes U(1)_Q}$ }
\def\SM{$\mathrm{SU(3)_c \otimes SU(2)_L \otimes U(1)_Y}$}

\def\3211{$\mathrm{SU(3) \otimes SU(2)_L \otimes U(1)_R \otimes U(1)_{B-L}}$ }
\def\321{$\mathrm{SU(3) \otimes SU(2) \otimes U(1)}$ }
\def\422{$\mathrm{SU(4) \otimes SU(2) \otimes SU(2)_R}$ }
\newcommand {\ignore}[1]{}

\newcommand{\sm}{{Standard Model }}

\def\SM{$\mathrm{ SU(3)_C \otimes SU(2)_L \otimes U(1)_Y }$}

\newcommand{\AddrAHEP}{%
  AHEP Group, Institut de F\'{i}sica Corpuscular --
  CSIC-Universitat de Val\`{e}ncia, Parc Cient\'ific de Paterna.\\
 C/ Catedr\'atico Jos\'e Beltr\'an, 2 E-46980 Paterna (Valencia) - SPAIN}
 
\newcommand{\iiserb}{Department of Physics, Indian Institute of Science Education and Research - Bhopal\\
  Bhopal Bypass Road, Bhauri, Bhopal, India}

\newcommand{\AddrFISTEO}{Departament de F\'{\i}sica Te\`{o}rica, Universitat de Val\`{e}ncia, 46100 Burjassot, Spain}

\begin{document}


\title{The Inverse Seesaw Family: Dirac And Majorana}

 \author{Salvador Centelles Chuli\'{a}}\email{salcen@ific.uv.es}
  \affiliation{\AddrAHEP}
   \author{Rahul Srivastava}\email{rahul@iiserb.ac.in}
    \affiliation{\iiserb}
     \author{Avelino Vicente}\email{avelino.vicente@ific.uv.es}
     \affiliation{\AddrAHEP}
     \affiliation{\AddrFISTEO}

 \begin{abstract}
   \vspace{1cm}

   After developing a general criterion for deciding which neutrino
   mass models belong to the category of inverse seesaw models, we
   apply it to obtain the Dirac analogue of the canonical Majorana
   inverse seesaw model. We then generalize the inverse seesaw
   model and obtain a class of inverse seesaw mechanisms both for
   Majorana and Dirac neutrinos. We further show that many of the
   models have double or multiple suppressions coming from tiny
   symmetry breaking ``$\mu$-terms''.  These models can be tested both
   in colliders and with the observation of lepton flavour violating
   processes.
    
 \end{abstract}

 \maketitle

\section{Introduction}
\label{sec:intro}

The discovery of neutrino oscillations provided us with one of the
first and clearest experimental hints of shortcomings in the \sm \, 
(SM). This is because the SM predicts neutrinos to be exactly massless
while neutrino oscillations irrefutably prove that at least two of the
three known neutrinos should carry mutually non-degenerate
masses~\cite{deSalas:2017kay,deSalas:2020pgw}.  However, neutrino
masses were theoretically anticipated by model builders much before
the experimental confirmation came. In fact, the so-called type-I
seesaw was developed in 1977~\cite{Minkowski:1977sc} a good 20+ years
before the first unambiguous experimental proof of neutrino
oscillations emerged. Ever since the first works, a plethora of
``neutrino mass models'' and ``neutrino mass generation mechanisms''
have been developed.~\footnote{These two terms will be more carefully
  defined in Section~\ref{sec:inv-seesaw}.}

The various neutrino mass generation models and mechanisms primarily
aim to generate non-zero neutrino masses as well as to provide an
explanation for their smallness with respect to the mass of the other
fermions. This is typically achieved through:
\begin{enumerate}[(a)]
 \item \textbf{Seesaw Mechanisms:} Neutrino masses are inversely proportional
   to a large
   scale~\cite{Minkowski:1977sc,Yanagida:1979as,Mohapatra:1979ia,GellMann:1980vs,Mohapatra:1980yp,Schechter:1980gr,Foot:1988aq,Ma:2014qra, Ma:2015mjd,Chulia:2016ngi}.
 \item \textbf{Loop Mechanisms:} Neutrino masses are generated as quantum
   corrections at loop
   level~\cite{Zee:1980ai,Cheng:1980qt,Zee:1985id,Babu:1988ki,Ma:2006km,Cai:2017jrq,Bonilla:2016diq,Bonilla:2018ynb}.
 \item \textbf{Naturalness Mechanisms:} Neutrino masses are directly
   proportional to a very small parameter~\cite{Mohapatra:1986bd,Akhmedov:1995vm} whose smallness is justified
   through 't Hooft naturalness criterion~\cite{tHooft:1979rat}.
 \item \textbf{Hybrid Mechanisms:} Neutrino masses are small due to some
   combination of the above three mechanisms.
\end{enumerate}

Before going on further, let us first briefly discuss the effective
operator approach to generate neutrino masses. Even though we will
focus in this work on completely renormalizable models, this will
serve as a guiding tool that will allow us to easily clasify
models. Since the nature of neutrinos is still unknown, we must
consider both possibilities of Dirac and Majorana neutrinos. In fact,
these operators are different for Majorana and Dirac neutrinos, as we
now proceed to discuss.

For {\bf Majorana neutrinos}, the effective operator behind neutrino
mass generation can generically be written
as~\cite{CentellesChulia:2018gwr,Anamiati:2018cuq}
\begin{equation} \label{eq:opM}
  \mathcal{L}_M = \frac{\mathcal{C}_M}{\Lambda^{m + n - 1}} \, \bar L^c \, L \, \Phi^{(m)} \, \sigma^{(n)}   + \hc \, ,
\end{equation}
where $\mathcal{C}_M$ is an effective coupling constant matrix, $L$ is
the usual SM lepton doublet and the generation indices are suppressed
for brevity. $\Lambda$ is the energy scale at which the new degrees of
freedom responsible for the generation of this effective operator
lie. Furthermore, $\Phi^{(m)}$ is a scalar operator with $m \geq 0$
$\rm SU(2)_L$ scalar doublets which need not be all of the same
type. Similarly, $\mathcal \sigma^{(n)}$ denotes a scalar operator
containing $n \geq 0$ (same or different types) scalar fields having
any $\rm SU(2)_L$ representation apart from the doublet
representation. Needless to say, none of the fields in $\Phi^{(m)} \,
\sigma^{(n)}$ that obtain a vacuum expectation value (VEV) should
carry nontrivial color or electric charges. Finally, the $\rm SU(2)_L$
representations of all the scalars should be such that the combination
$\Phi^{(m)} \, \sigma^{(n)}$ transforms either as a triplet or singlet
under $\rm SU(2)_L$. This will then ensure that the effective operator
in Eq.~\eqref{eq:opM} is a singlet under the \SM \, gauge symmetry. We
note that $m = 2, n = 0$, with $\Phi \equiv H$ the SM Higgs doublet,
would lead to the well-known Weinberg operator~\cite{Weinberg:1979sa},
without additional scalar VEV insertions. Another popular example is
obtained with $m=2$, $n=1$, $\Phi = H$ and $\sigma = \chi$, where
  $\chi$ is a singlet scalar field. This would be the
effective operator of the majoron
model~\cite{Chikashige:1980ui,Schechter:1981cv}. Operators including
$\Phi^{(2)} = H_a H_b$, with $H_a$ and $H_b$ two different Higgs
doublets, are induced in the context of the Two-Higgs-doublet model
framework, for instance in supersymmetric models~\cite{Krauss:2011ur}.

In case of {\bf Dirac neutrinos}, the SM particle content must
necessarily be extended to include right-handed neutrinos $\nu_R$,
singlets under the SM gauge group. Being SM singlets, the number of
right-handed neutrinos is unconstrainted by theory. However, at least
three right handed neutrinos are needed to generate sequential Dirac
masses for all the three neutrino flavours. The effective operator
leading to neutrino masses in this scenario can be written
as~\cite{CentellesChulia:2018gwr, CentellesChulia:2018bkz}
\begin{equation} \label{eq:opD}
  \mathcal{L}_D = \frac{\mathcal{C}_D}{\Lambda^{m + n - 1}} \, \bar L \, \nu_R \, \Phi^{(m)}\, \mathcal \sigma^{(n)} + \hc \, ,
\end{equation}
where $\mathcal{C}_D$ is an effective coupling constant matrix and we
have followed the same notation as in Eq.~\eqref{eq:opM}, again
suppressing generation indices. The usual conditions mentioned in the
Majorana case apply here too, except that the scalar combination
$\Phi^{(m)} \, \sigma^{(n)}$ should now transform as an $\rm SU(2)_L$
doublet to ensure that the operator in Eq.~\eqref{eq:opD} is a singlet
under \SM. Notice that the simplest case will be $m=1, n=0$ with $\Phi
\equiv \widetilde{H}$, where we have defined $\widetilde{H} = i \tau_2
H^\ast$, with $\tau_2$ the second Pauli matrix. Another simple
realization would be $m = 1, n=1$ with $\Phi \equiv \widetilde{H}$ and
$\sigma \equiv \chi$ being an \SM \,
singlet~\cite{Chulia:2016ngi}. Other realizations are also possible,
see~\cite{Yao:2017vtm, CentellesChulia:2018gwr, Yao:2018ekp,
  CentellesChulia:2018bkz,CentellesChulia:2019xky}.

In this work we aim to look in detail at one of the most popular
naturalness mechanisms, the inverse seesaw
mechanism~\cite{Mohapatra:1986bd}. We will do so for both Dirac and
Majorana neutrinos and demonstrate the various possibilities for both
scenarios by constructing several explicit models.  Special attention
will be given to the Dirac realization of the mechanism, which has
been only briefly discussed in the literature~\cite{Borah:2017dmk}. We
will show that specific models for Dirac neutrinos can be built with
the same defining properties that characterize the well-known Majorana
inverse seesaw. Most of the models discussed here are, to the best of our
knowledge, constructed for the first time in this work.

The rest of the manuscript is organized as follows. We begin by
defining the key features of the inverse seesaw mechanism in
Section~\ref{sec:inv-seesaw}. Then, in order to fix notations and
conventions, we describe the well-known Majorana inverse seesaw in
Section~\ref{sec:Majorana}, in which the initial $\rm U(1)_{B-L}$ symmetry
gets broken to a residual $\mathbb{Z}_2$. Then we proceed to present
the Dirac versions of the inverse seesaw in
Section~\ref{sec:Dirac}. Sections~\ref{sec:generalized1} and
\ref{sec:generalized2} discuss various generalizations of the minimal
inverse seesaw setups presented in the previous Sections, both for
Dirac and Majorana neutrinos. In Section~\ref{sec:generalized1} we
explore versions of the inverse seesaw containing additional
representations of the SM gauge group, while
Section~\ref{sec:generalized2} considers extensions with additional
fermionic states that lead to further suppressions of the resulting
light neutrino masses. Finally, we summarize our results and draw
conclusions in Section~\ref{sec:summary}.

\section{The inverse seesaw framework}
\label{sec:inv-seesaw}

The inverse seesaw is a popular approach for the generation of
neutrino masses with the mediator masses potentially being close to
the electroweak scale. It is characterized by the presence of a small
mass parameter, generally denoted by $\mu$, which follows the
hierarchy of scales
\begin{equation} \label{eq:hierar}
 \mu \ll v \ll \Lambda \, ,  
\end{equation}
with $v$ the Higgs VEV that sets the electroweak scale and $\Lambda$
the neutrino mass generation scale, determined by the masses of the
seesaw mediators. The $\mu$-parameter suppresses neutrino masses as
$m_\nu \propto \mu$, allowing one to reproduce the observed neutrino
masses and mixing angles with large Yukawa couplings and light seesaw
mediators. This usually leads to a richer phenomenology compared to
the standard high-energy seesaw scenario.

In this paper we will explore the various types of inverse seesaws
possible for both Dirac and Majorana neutrinos. To do this let us
begin with an attempt to first precisely define what we mean by mass models 
and mass mechanisms. \\ \\
\textbf{Neutrino Mass Generating Models:} A proper neutrino mass model
should be capable of generating neutrino masses and should be
renormalizable. It is also highly desirable, though not essential,
that the mass model also provides an ``explanation'' for the non-zero
yet so tiny masses of neutrinos when compared to masses of all the
other fermions in the SM.  \\
\textbf{Neutrino Mass Generation Mechanisms:} A mechanism is a class
of models which generates the neutrino masses in the same or very closely
related ways. For example, various variants of the canonical type-I
seesaw model can be clubbed together as type-I seesaw mechanism.
\\ \\
Given the above definition of the mass generation models and
mechanisms we can now define the criterion to determine which models
can be classified as belonging to the inverse seesaw mechanism:
\begin{enumerate}
\item \textbf{Presence of a Small Symmetry Breaking Parameter :} The
  first and foremost condition for a model to be classified as an
  inverse seesaw model is the requirement of a ``small'' symmetry
  breaking ``$\mu$-parameter''.  The $\mu$-parameter has to be such
  that the limit $\mu \to 0$ enhances the symmetry of the
  Lagrangian. This crucial feature implies that in the absence of the
  $\mu$-parameter, the model would have a conserved symmetry group
  $\mathcal{G}$, which gets broken by $\mu \ne 0$ as
\begin{equation}
\mathcal{G} \, \xrightarrow{\hspace*{0.4cm} \mu \hspace*{0.4cm}} \, \mathcal{G}^\prime  \, .
\label{eq:mu-term}
\end{equation}
Here $\mathcal{G}^\prime \supset \mathcal{G}$ is a residual
symmetry.\footnote{It can happen that the $\mu$-parameter completely
  breaks the symmetry group $\mathcal{G}$. In such a case
  $\mathcal{G}' \equiv \mathcal{I}$ i.e. the trivial Identity Group.}
Therefore, the limit $\mu \to 0$ enhances the symmetry of the model,
making it natural in the sense of 't Hooft~\cite{tHooft:1979rat} and
protecting the small value of $\mu$ from quantum corrections. Note
that here smallness\footnote{We leave the ``How small should be
  considered small?'' question to the model creator's taste.} of the
$\mu$-parameter is with respect to other parameters in the model under
consideration. When the dimensions of the other parameters in the
model do no match the dimensions of the $\mu$-parameter, the other
parameters should be correctly normalized before making the
comparison.

\item \textbf{$\boldsymbol{\mu}$-parameter from Explicit/Spontaneous
  Symmetry Breaking:} The $\mu$-parameter can either be an explicit
  symmetry breaking term or a spontaneously induced symmetry breaking
  term. However, to classify as a genuine inverse seesaw, the $\mu$-parameter 
  should be a ``soft term''. In particular, this means that
  if the $\mu$-parameter is an explicit symmetry breaking term, then
  it should have a positive mass dimension.

\item \textbf{Neutrino Mass Dependence on
  $\boldsymbol{\mu}$-parameter:} The neutrino mass at leading order
  must be directly proportional to the $\mu$-parameter.

\item \textbf{Extended Fermionic Sector:} A genuine inverse seesaw
  model should always have an extended fermionic sector directly
  participating in the neutrino mass mechanism. This means fermions
  beyond the fermionic content of the SM should be involved in
  neutrino mass generation.

\item \textbf{The $\boldsymbol{\mu}$-parameter need not be unique:} In
  cases where there are different $\mu_i$-parameters, all should be soft
  and in the limit of $\mu_i \to 0\, \, \forall \, \, i $, the symmetry
  of Lagrangian should be enhanced. Also, at least one $\mu_i$-parameter
  should be directly involved in the neutrino mass generation
  mechanism. Furthermore, all the $\mu_i$-parameters directly involved in
  neutrino mass generation should satisfy all the other conditions
  listed above.

\end{enumerate}

An example of a model which satisfies all these features is the
canonical Majorana inverse seesaw model~\cite{Mohapatra:1986bd}. The
SM field inventory is extended to include a Vector Like (VL) fermion
transforming as a singlet under the gauge group. The explicit Majorana
mass term (a soft term) for this new fermion will break lepton number
in two units explicitly and thus its smallness is protected by a
symmetry. In this notation, $\mathcal{G} = \rm{U(1)_L}$ while
$\mathcal{G}^\prime = \mathbb{Z}_2$ and $\mu$ is the explicit
Majorana mass term. More details are given in
Section~\ref{sec:Majorana}.

Let us emphasize again that the $\mu$-parameter can be explicitly
introduced in the Lagrangian, as a symmetry-breaking mass term, or
spontaneously generated by the VEV of a scalar. In the rest of the
paper we will concentrate on the latter case. This is particularly
convenient for our discussion, since the identification of the broken
symmetry becomes more transparent. Furthermore, the smallness of the
$\mu$-parameter can be more easily justified in extended models that
generate it spontaneously. We note, however, that scenarios with an
explicit $\mu$-term would lead to analogous conclusions, just
replacing a VEV by a bare mass term.~\footnote{Note that this analogy
  is only true if the $\mu$-term does not break any of the SM gauge
  symmetries. Otherwise, explicit violation is forbidden while
  scenarios with spontaneous violation are in principle allowed,
  provided $\mu \ll v$, as generally assumed in the inverse seesaw
  setup. Therefore, the spontaneously broken scenario is in a sense
  more general than the explicitly broken one. Of course, electric
  charge and color should remain as conserved charges in either case.}

\section{Warm up: Canonical Majorana inverse seesaw}
\label{sec:Majorana}

As a warm up, we will start by fleshing out the well-known case of the
canonical Majorana inverse seesaw, in which the $\rm U(1)_{L}$, or
equivalently $\rm U(1)_{B -L}$, symmetry is broken to a residual
$\mathbb{Z}_2$ subgroup~\cite{Mohapatra:1986bd}. While this symmetry
breaking is typically done explicitly, here keeping in mind the ease
of generalization and the clarity it offers regarding the residual
subgroup, we will construct a fully consistent model in which the $\rm
U(1)_{B -L}$ symmetry is broken spontaneously.  Although the model
shown here is of course not new, it will allow us to set up the
notation and conventions.

\begin{table}[h]
\begin{center}
\begin{tabular}{| c | c | c   || c | c | c |}
  \hline 
   Fields            &     \hspace{0.2cm} $\rm SU(2)_{L} \otimes U(1)_Y$            
&\hspace{0.05cm}$\rm U(1)_{B-L}$ \hspace{0.05cm}$\to$\hspace{0.05cm} $\mathbb{Z}_2$\hspace{0.05cm} &   Fields            &     \hspace{0.2cm} $\rm SU(2)_L \otimes U(1)_Y$           
&\hspace{0.05cm}$\rm U(1)_{B-L}$ \hspace{0.05cm}$\to$\hspace{0.05cm} $\mathbb{Z}_2$\hspace{0.05cm}                            \\
\hline \hline
   $L$        	  &   ($\mathbf{2}, {-1/2}$)        &    $-1 \, \, \to \, \, -1$   &
                      &                                 &          	    	        \\	
   $N$              &   ($\mathbf{1}, {0}$)           &    $1 \, \, \to \, \, -1$   &
   $S$              &   ($\mathbf{1}, {0}$)           &    $-1 \, \, \to \, \, -1$  \\
\hline \hline
   $H$  	          &  ($\mathbf{2}, {1/2})$          &    $0 \, \, \to \, \, 1$     &
   $\chi$ 	          &  ($\mathbf{1}, 0)$              &   $2 \, \, \to \, \, 1$  \\
    \hline
  \end{tabular}
\end{center}
\caption{Particle content of the model. The $\rm U(1)_{B-L}$ symmetry gets broken into the residual $\mathbb{Z}_2$ after $\chi$ gets a VEV. All fields are taken to be left-handed. Note that $N$ and $S$ form a VL pair of fermions.
\label{tab:MajoranaZ2}}
\end{table}

The symmetries and field inventory of the model are shown in
Tab.~\ref{tab:MajoranaZ2}. In what concerns the number of generations
of the new fields, it is common to assume $3$ copies for each species,
although more minimal options
exist~\cite{Abada:2014vea,Rojas:2019llr}. Under the assumption of a
conserved $\rm U(1)_{B-L}$ symmetry, one can write the Lagrangian
terms
\begin{equation}
\label{eq:majoranaz2lag1}
\mathcal{L}_{\rm Maj} \, = \, Y \, \bar{L}^c \, \widetilde H \, N \, + \, \lambda \, \bar{S}^c \,  \chi \, S \, + \, M \, \bar{S}^c \, N \, + \hc \, .
\end{equation}
where $\widetilde{H} = i \tau_2 H^*$, with $\tau_2$ the second Pauli
matrix. The new fermions $N$ and $S$ and the scalar $\chi$ are all \sm \, 
gauge singlets. However, they all carry $\rm B - L$ charges given by
$N = 1$, $S = -1$ and $\chi = 2$.  Note that, unless stated otherwise,
the generation indices in Eq.~\eqref{eq:majoranaz2lag1} as well as
throughout this paper are suppressed for brevity. The scalars $H$ and
$\chi$ obtain VEVs given by
\begin{equation}
\langle H \rangle = v \quad , \quad \langle \chi \rangle = u \, .
\end{equation}
Here $v$ is the usual SM Higgs VEV, responsible for the breaking of
the electroweak symmetry, while $u$ breaks $\rm B-L$ in two units,
leaving a residual $\mathbb{Z}_2$ symmetry. 
In fact, the $\chi$ VEV induces a Majorana mass term for the $S$ fermion,
\begin{equation}
\mu = \lambda \, u \, .
\end{equation}
This is the usual $\mu$-parameter in the standard inverse seesaw
model which, in the literature, is often put as an explicit
symmetry breaking term.  Since $\mu$ is a symmetry-breaking term and
in the limit $\mu \to 0$ the Lagrangian has the enhanced $\rm
U(1)_{B-L}$ symmetry, $\mu$ can be naturally small. This implies that
in the spontaneous symmetry breaking version currently under
consideration, one naturally has $u \ll v$ for $\lambda \approx
\mathcal{O}(1)$.

We would like to emphasize that the t'Hooft naturalness condition
applied here, merely states that if $\mu$ is small then its smallness
will be protected against quantum corrections i.e. $\mu$ will not
receive any large quantum corrections. However, the naturalness
condition does not explain why $\mu$ should be small in the first
place. Of course, one can indeed ask why $\mu$ should be small. One possible
answer is that, maybe the smallness of $\mu$ is owing to the fact that
the origin of such symmetry breaking terms lies in a bigger theory. In
such a bigger theory, its smallness is due to suppression by some
large scale or, alternatively, because it is generated at loop level,
see~\cite{Bazzocchi:2009kc, Bazzocchi:2010dt, Rojas:2019llr} for
examples. However, in this work we do not attempt to address the cause
for initial smallness of $\mu$ in any detail and will simply assume
that $\mu$ is small to begin with. Then in such a case the t'Hooft
naturalness criterion will ensure that it remains small even after
quantum corrections.

\begin{figure}[t!]
\centering
\includegraphics[scale=0.7]{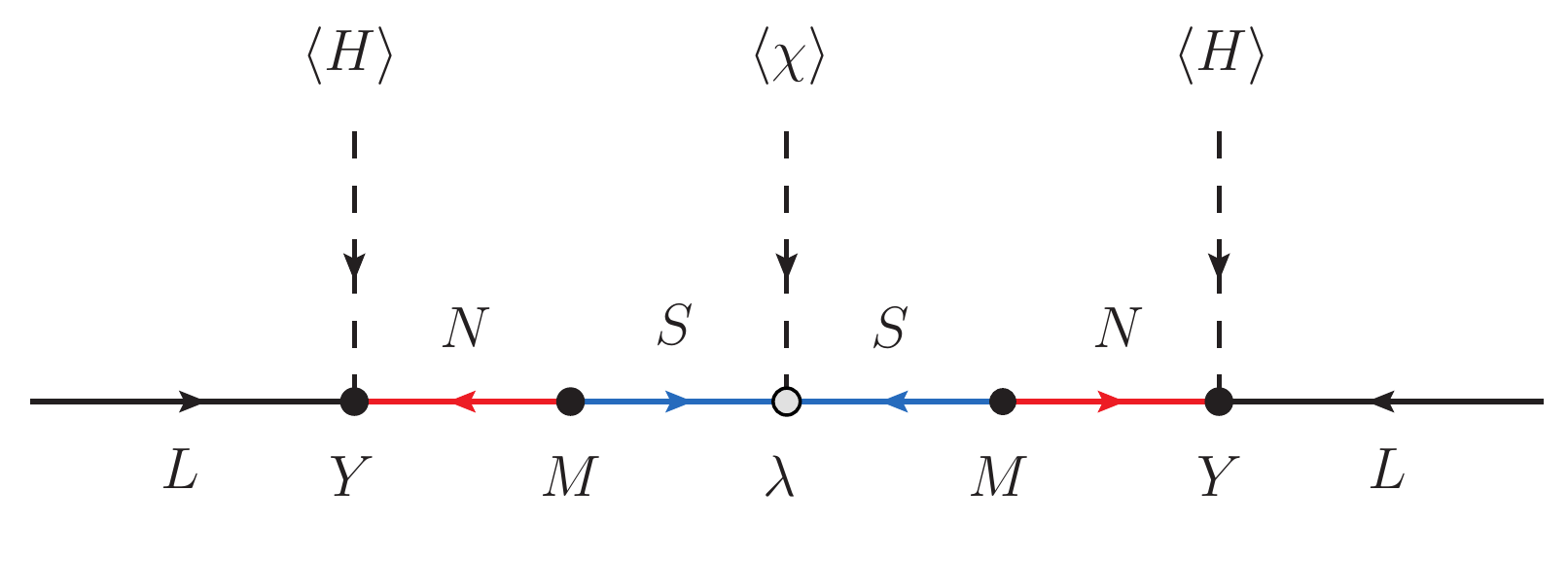}
\caption{Neutrino mass generation in the standard Majorana inverse
  seesaw. The $\mu$-term can be either spontaneously generated by the
  VEV of the scalar $\chi$ or can be explicitly added as a soft $\rm B-L$ breaking term.
  \label{fig:MajoranaZ2}}
\end{figure}

Coming back to the canonical inverse seesaw model, we note that the
symmetries of the model will \textit{always} allow the Yukawa term
\begin{equation}
\label{eq:majoranaz2lag2}
\mathcal{L}_{\rm Maj}^\prime \, = \, \lambda^\prime \, \bar{N}^c \, \chi^* \, N \, + \hc
\end{equation}
which preserves the $\rm U(1)_{B-L}$ symmetry. After symmetry
breaking, this new piece generates a second Majorana mass term, in
this case for the $N$ fermion,
\begin{equation}
\mu^\prime = \lambda^\prime \, u \, .
\end{equation}
Eqs.~\eqref{eq:majoranaz2lag1} and \eqref{eq:majoranaz2lag2} lead to
the following Majorana mass term after symmetry breaking
\begin{equation}
 \mathcal{L}_{m} \, = \, \left ( \begin{matrix}
          \bar{\nu}^c & \bar{N}^c & \bar{S}^c
         \end{matrix}
 \right )
 \left ( \begin{matrix}
          0      & Y \, v  & 0 \\
          Y^T \, v  & \mu^\prime   & M^T \\
          0      & M         & \mu          
          \end{matrix}
 \right )
 \left ( \begin{matrix}
          \nu \\
          N \\
          S
         \end{matrix}
 \right ) \, .
\end{equation}     
If the model parameters follow the inverse seesaw hierarchy of
Eq.~\eqref{eq:hierar}, or equivalently,
\begin{equation} \label{eq:seesaw}
\mu , \mu^\prime \ll Y \, v \ll M \, ,
\end{equation}
then the light neutrino mass matrix can be obtained in seesaw
approximation as
\begin{eqnarray}
 m_\nu &=&  \left ( \begin{matrix}
          Y  v & 0
         \end{matrix}
 \right )
 \left ( \begin{matrix}

          \mu^\prime      & M^T \\
           M         & \mu          
          \end{matrix}
 \right )^{-1}
 \left ( \begin{matrix}
          Y^T  v \\
          0
         \end{matrix}
 \right ) \, .
\end{eqnarray}      
Assuming one generation for the time being, this leads to the light
neutrino mass formula
\begin{equation}
 m_\nu = Y^2 \, \frac{v^2 \mu}{\mu \, \mu^\prime - M^2} \, .
 \label{eq:gen-class-maj-inv-mass}
\end{equation}
The first term in the denominator of
Eq.~\eqref{eq:gen-class-maj-inv-mass} is negligible under the seesaw
hierarchy in Eq.~\eqref{eq:seesaw}, and one can finally approximate
\begin{equation} \label{eq:MajoranaMass}
 m_\nu = - Y^2 \, \frac{v^2 \, \mu}{M^2} \, ,
\end{equation}
and obtain the usual inverse seesaw formula, diagramatically
represented in Fig.~\ref{fig:MajoranaZ2}.

The inverse seesaw mechanism has several appealing features. First of
all, we note that the resulting neutrino mass formula turns out to be
proportional to the small $\mu$-parameter. This allows one to obtain
light neutrino masses of the order of $0.1$ eV with TeV-scale seesaw
mediators and $\mathcal{O}(1)$ Yukawa couplings, hence leading to a
much richer phenomenology compared to the usual high-scale seesaw
scenarios. Furthermore, we note that the $\mu$-parameter is protected
by the $\rm B-L$ symmetry, since the limit $\mu \to 0$ restores $\rm
U(1)_{B-L}$. This makes its smallness perfectly natural in the sense
of 't Hooft. Finally, we also note that in this construction the $\rm
U(1)_{B-L}$ symmetry (or equivalently $\rm U(1)_{L}$) remains
anomalous and therefore cannot be gauged. In this case they must be
global symmetries, and their spontaneous breaking (by $u \ne 0$) would
lead to the appearance of a Goldstone boson, the majoron.

On a side note, the reader should keep in mind that from the diagram
depicted in Fig.~\ref{fig:MajoranaZ2} alone, one cannot distinguish between
the simplest scenario in which $\rm{U(1)_{B-L}} \rightarrow
\mathbb{Z}_2$ and more involved situations with $\rm{U(1)_{B-L}}
\rightarrow \mathbb{Z}_{2n}$. Note that, when neutrinos are Majorana
fermions, the residual transformation of the light neutrinos will
always be $\nu \sim z^n = -1$ where $z = e^{i \pi/n}$. The
difference between $\mathbb{Z}_2$ and a more complicated
$\mathbb{Z}_{2n}$ cannot be seen in the neutrino mass generation but
will generate differences in the scalar sector. A detailed discussion
between the differences of these types of models shall be found
elsewhere since it would be beyond the scope of this work.

\section{The simplest Dirac inverse seesaw}
\label{sec:Dirac}

In this section we aim to develop the inverse seesaw model for Dirac neutrinos. 
However, before going into details of the model, let us switch over to the ``chiral notation'' first. Thus, from now on we give the same field name to fermions who will ultimately form a VL pair. The left- and right-handed fields are distinguished with a subscript $L$ or $R$.
For example, the fields of Section~\ref{sec:Majorana} get renamed as
$N \to N_L$ and $S^c \to N_R$ in the chiral notation. This notation
change is done to facilate the user to easily identify the fields
which will ultimately form either a Dirac or pseudo-Dirac pair. It
will become especially useful when working with Dirac neutrinos.

Furthermore, as mentioned already, throughout this work we consider
symmetries to be always broken spontaneously. This choice is taken
because we find that the symmetry transformations of the fields before
and after breaking, as well as, the nature of the residual symmetry
that is left, are more transparent when the symmetry is spontaneously
broken. Also, as argued before, the spontaneous version is completely
general and for any model with an explicit symmetry breaking
$\mu$-term, its spontaneous symmetry breaking analogue can be always
constructed. It has the added advantage that in cases where the
symmetry under consideration is a gauge symmetry, only the spontaneous
breaking of the symmetry is allowed and hence only spontaneously
broken models are mathematically consistent.

Finally, before moving on we want to reiterate that in this work we do not attempt to answer the questions: 
\begin{itemize}
 \item How and why the $\mu$-term should be small in the first place?
 \item How small a $\mu$-term should be considered small?
\end{itemize}
These are important questions and are left to be addressed by the
creators of a given model. We are merely assuming that, to begin with,
the $\mu$-term is small. Moreover, since in the limit $\mu \to 0$ the
symmetry of the system is enhanced, therefore following t'Hooft's
arguments, such a small $\mu$-term will be protected against large
quantum corrections and will remain small. In the spontaneous symmetry
breaking versions that we consider, this means that the VEV of the
scalar which leads to the $\mu$-term will be much smaller than the
electroweak VEV.

Coming back to the Dirac inverse seesaw model, note that for neutrinos
to be Dirac particles, some unbroken symmetry should forbid the
appearance of Majorona mass terms. This symmetry can very well be the
residual symmetry $\mathcal{G}'$ left unbroken when the $\mu$-term
breaks the bigger symmetry $\mathcal{G}$, see
Eq.~\eqref{eq:mu-term}. Here we will take $\mathcal{G} = \rm
U(1)_{B-L}$ and $\mathcal{G}'$ can be any of its $\mathbb{Z}_n$; $n >
2$ subgroups~\cite{Hirsch:2017col}. In this section we take the
simplest possibility of $\mathcal{G}' = \mathbb{Z}_3$.

Symmetries play an even more central role in Dirac inverse seesaw
constructs. First, a symmetry is required to ensure the Dirac nature
of neutrinos i.e. to forbid Majorana mass terms for them.
Second, a symmetry is also necessary to forbid the tree-level term
$\bar{L} \widetilde H \nu_R$, which if present would imply tiny Yukawa
couplings.
Third, one must resort to a symmetry breaking argument to make the
smallness of the $\mu$-parameter natural.
One can accomplish these tasks by using different symmetries for each
task. However, as we discuss now, the $\rm U(1)_{B-L}$ symmetry and
its residual $\mathbb{Z}_3$ subgroup are enough to play all these
roles.

We are choosing here the $\rm U(1)_{B-L}$ symmetry because it can be
made anomaly free by adding right-handed neutrinos with appropriate
$\rm B-L$ charges. This can be especially important if we were to
gauge the symmetry, as is often done. The usual solution to make $\rm
U(1)_{B-L}$ anomaly free is to add three right-handed neutrinos
$\nu_R$ with $\rm B-L$ charges $(-1, -1, -1)$. However, an exotic
choice of $\rm B-L$ charges for the right-handed neutrinos can fulfill
all these conditions, the so-called \textit{445 chiral
  solution}~\cite{Montero:2007cd,Ma:2014qra,Ma:2015mjd}. In this case,
the right-handed neutrinos carry $(-4, -4, 5)$ charges under $\rm
U(1)_{B-L}$. Being anomaly free, the $\rm U(1)_{B-L}$ symmetry can
also be gauged, leading to a richer phenomenology.  Throughout this
work we will mainly use this solution in all the Dirac models that we
will construct. However, let us mention that this is not the only
possible symmetry solution~\cite{Bonilla:2018ynb} but just a
particularly elegant one.

\subsection*{Dirac Inverse Seesaw}
\label{subsec:dirac-inv}

\begin{table} [htb!]
\begin{center}
\begin{tabular}{| c || c | c | c  || c | c | c |}
  \hline 
&   Fields            &     \hspace{0.2cm} $\rm SU(2)_L \otimes U(1)_Y$                         
&\hspace{.05cm} $\rm U(1)_{B-L}$ \hspace{.05cm}$\to$\hspace{.05cm} $\mathbb{Z}_3$ \hspace{.05cm}                         &   Fields            &     \hspace{0.2cm} $\rm SU(2)_L \otimes U(1)_Y$                         
&\hspace{.05cm} $\rm U(1)_{B-L}$ \hspace{.05cm}$\to$\hspace{.05cm} $\mathbb{Z}_3$ \hspace{.05cm}
\\
\hline \hline
\multirow{4}{*}{ \begin{turn}{90} \hspace{0.9cm} \tiny{Fermions} \end{turn} }
&   $L_i$        	  &   ($\mathbf{2}, {-1/2}$)       &    $-1  \,\, \to \,\,  \omega^2$  
&   $\nu_R$      	  &   ($\mathbf{1}, {0}$)          &    $(-4, -4, 5)  \,\, \to \,\,  \omega^2$ \\	
&   $N_L$        	  &   ($\mathbf{1}, {0}$)          &    $-1  \,\, \to \,\,  \omega^2$
&   $N_R$        	  &   ($\mathbf{1}, {0}$)          &    $-1  \,\, \to \,\,  \omega^2$  \\	
\hline \hline 
\multirow{4}{*}{ \begin{turn}{90} \hspace{1.4cm}  \tiny{Scalars}\end{turn} }
&  \rule{0pt}{4ex} $H$     &   ($\mathbf{2}, {1/2}$)       &    $0  \,\, \to \,\,  \omega^0$   	      
&   $\chi$                 &   ($\mathbf{1}, {0}$)         &    $3  \,\, \to \,\,  \omega^0$  \rule[-3ex]{0pt}{0pt}   	       \\	
    \hline
  \end{tabular}
\end{center}
\caption{Particle content of the minimal model implementing the Dirac
  inverse seesaw.  All quarks transform as $1/3$ ($\omega^1$) under
  $\rm U(1)_{B-L}$ ($\mathbb{Z}_3$), while their \SM \, charges are
  identical to those in the SM. Here $\omega = e^{2\pi i/3}$ is the
  cube root of unity with $\omega^3 = 1$. Moreover, with this choice
  of charges the $\rm U(1)_{B-L}$ symmetry is anomaly free.
 \label{tab:minimaldirac}}
\end{table}

We begin the discussion on Dirac versions of the inverse seesaw
mechanism with a very minimal realization.  In this simple case, the
leading effective operator for neutrino masses is $\bar{L}
\widetilde{H} \chi \nu_R$. This corresponds to the operator in
Eq.~\eqref{eq:opD} with $\Phi \equiv \widetilde{H}$ where $H$ is the
Higgs doublet and $\sigma \equiv \chi$; $\chi$ being an \SM \,
singlet.
In the full ultraviolet complete theory, the particle content and
symmetry transformations are shown in Tab.~\ref{tab:minimaldirac},
while the relevant Lagrangian terms for the generation of neutrino
masses are given by
\begin{equation}
 \mathcal{L}_{\rm Min} = Y \, \bar{L} \widetilde{H} N_R \, + \, \lambda \, \bar{N}_L \chi \nu_R \, + \, M \, \bar{N}_L N_R \, + \hc \, .
\end{equation}
The scalar acquire VEVs  
\begin{equation}
\langle H \rangle = v \quad , \quad \langle \chi \rangle = u \, .
\end{equation}
and break the electroweak and $\rm U(1)_{B-L}$ symmetries,
respectively.  The VEV of $\chi$ induces the small symmetry breaking
$\mu$-term.
\begin{equation}
\mu = \lambda \, u \, .
\end{equation}
Also, note that the VEV of the singlet scalar $\chi$ breaks $\rm
U(1)_{B-L}$ in three units, leaving a residual
$\mathbb{Z}_3$ symmetry under which all scalars transform trivially
while all fermions (except quarks) transform as $\omega^2$, with
$\omega = e^{2 i \pi/3};\, \omega^3 = 1$ being the cube root of
unity. This symmetry forbids all Majorana terms and therefore protects
the Diracness of light neutrinos. The $H$ and $\chi$ VEVs also induce
\textit{Dirac} masses proportional to the $Y$ and $\lambda$ Yukawa
couplings.
These, in the Dirac basis $\left(\bar{\nu}_L \, \bar{N}_L \right)$ and
$\left(\nu_R \, N_R \right)^T$, gives rise to the mass matrix
\begin{equation}
 \mathcal{M}  =  
 \left ( \begin{matrix}
          0          & Y \, v\\
          \mu   & M
          \end{matrix}
 \right ) \, 
 \label{eq:min-dirac-mass}
\end{equation}
where, as mentioned before, $\mu = \lambda\, u$ will be naturally small as
it is the $\rm U(1)_{B-L}$ symmetry breaking term.

\begin{figure}[t!]
\centering
\includegraphics[scale=0.7]{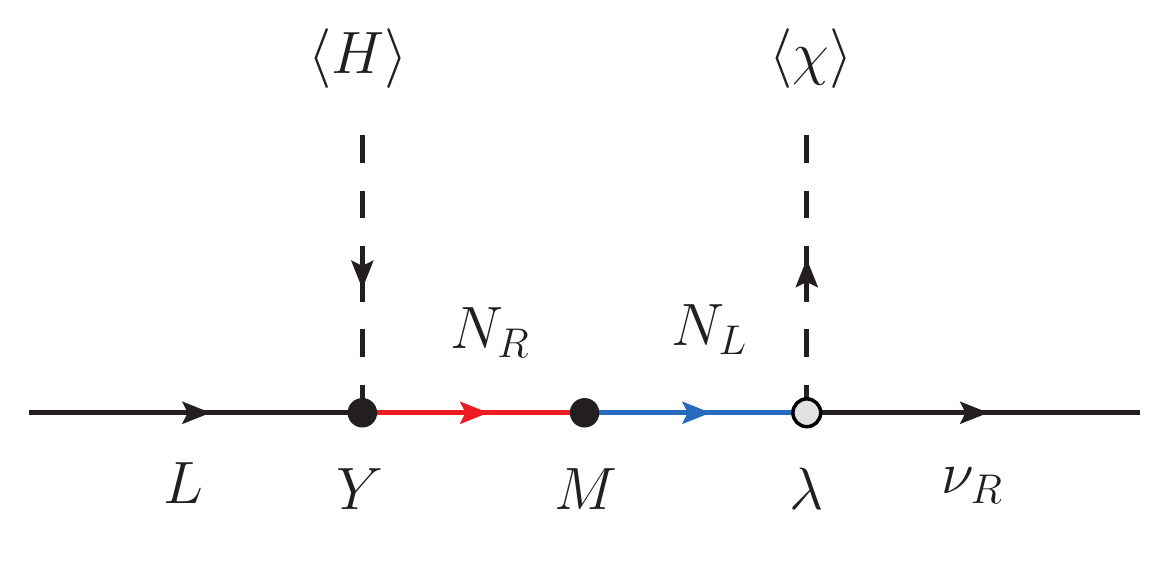}
\caption{Neutrino mass generation in the minimal Dirac inverse
  seesaw.
\label{fig:MinDirac}}
\end{figure}

Thanks to the residual $\mathbb{Z}_3$ symmetry, the neutrinos are Dirac
particles whose masses in the inverse seesaw limit $M \gg Y v \gg
\mu$ are given by
\begin{equation} \label{eq:MinDiracMass}
 m_\nu =   Y  \,v \, \frac{\mu}{M} \, ,
\end{equation}
as diagramatically shown in Fig.~\ref{fig:MinDirac}.

Since in the limit $\mu \to 0$ the $\rm U(1)_{B-L}$ symmetry is restored, the smallness of the $\mu$-term is protected. Its smallness will not be altered by higher order corrections, and therefore is perfectly natural. 
For example, if we take a small $\mu \sim 10$ eV, we can obtain $m_\nu \sim 0.1$ eV for $Y \sim 0.1$ and $M \sim 1$ TeV. 
We should also point out that the mass matrix in
Eq.~\eqref{eq:min-dirac-mass} is similar to the mass matrix one
obtains in the Dirac type-I
seesaw~\cite{Ma:2014qra,Ma:2015mjd,Chulia:2016ngi,CentellesChulia:2018gwr}. However,
in the type-I seesaw case, the off-diagonal terms in
Eq.~\eqref{eq:min-dirac-mass} are taken to be comparable to each other
i.e. $Y\, v \approx \lambda \, u \ll M$. Thus, for Dirac neutrinos, the
minimal inverse seesaw and the type-I seesaw are just two limits of
the same mass matrix.

Some final comments are in order. First, since the $\rm U(1)_{B-L}$
symmetry is anomaly free, it can be a gauge symmetry. This eliminates
the Goldstone boson associated to its breaking and leads to a much
richer phenomenology.
Note that the canonical Majorana inverse seesaw discussed in
Section~\ref{sec:Majorana} is not anomaly free and hence cannot be
gauged.
Second, the Dirac inverse seesaw model is also relatively simple in
terms of new fields added to the theory. In addition to the \sm \, 
particles, we have just added the right-handed neutrinos $\nu_R$, a
VL fermionic pair $N_L$ and $N_R$ and an extra singlet scalar
$\chi$, which is needed only if the spontaneous symmetry breaking is
desired.


\section{Generalizing the Inverse Seesaw - I : Multiplets}
\label{sec:generalized1}


The canonical Majorana inverse seesaw and its Dirac analogue discussed
in the previous sections are the simplest possibilities to implement
the inverse seesaw mechanism. However, inverse seesaw as an idea is
much more general and can be implemented in many different ways. It is
the aim of this and Section~\ref{sec:generalized2} to explore the
various ways in which one can generalize it. As we show, there are
several directions in which both the Dirac and Majorana inverse
seesaws can be generalized. Many of these generalized models contain
exotic fermions and scalars which will have very unique signatures in
experiments, which we intend to explore in followup work.

In this section we restrict ourselves only to generalizing the $\rm
SU(2)_L \times U(1)_Y$ multiplets that can lead to the inverse seesaw,
both for Majorana and Dirac neutrinos. We will also show a few
examples of each kind. We emphasize that these generalized versions
may have a richer collider phenomenology than their simpler cousins.

\subsection{Generalized Majorana inverse seesaw}
\label{sec:generalizedmajorana}

We consider a generalization of the Majorana inverse seesaw that uses
the same number of scalars and fermions as in
Tab.~\ref{tab:MajoranaZ2} and Fig.~\ref{fig:MajoranaZ2}, but allows
for other representations under $\rm SU(2)_L \times U(1)_Y$. The
Lagrangian will be formally equivalent to the one shown in
\ref{sec:Majorana} just replacing $H \rightarrow \phi$, $\chi \rightarrow \varphi$, $N \rightarrow N_L$ and $S^c \rightarrow N_R$, but with different multiplets of $\rm
SU(2)_L$. There are many such generalizations possible and before we
embark on their discussion, we need to streamline the notation such
that the same notation can be easily applicable to Dirac as well as
Majorana cases and to various possible generalizations. This
generalized notation will be used throughout the rest of the paper.
\begin{itemize}

\item The two new type of fermions (called $N,S$ in
  Section~\ref{sec:Majorana}) which will ultimately form a VL
  pair are henceforth called by $N_L$ and $N_R$. Note that they have
  the same $\rm SU(2)_L \otimes U(1)_Y$ representation and $\rm B-L$ charges.
  Henceforth, we will denote the $\rm SU(2)_L$ and
  $\rm U(1)_Y$ charges of a given particle as $(\textbf{n}', Y')$; $\textbf{n}'$ being the
  dimensionality of the $\rm SU(2)_L$ multiplet and $Y'$ being its
  hypercharge. In this notation, the two new fermions of
  Section~\ref{sec:Majorana} will both have their $\rm SU(2)_L \otimes
  U(1)_Y$ charges given by $(1, 0)$.
\item We denote the scalar which couples to $L$ and $N_R$ as
  $\phi$. This scalar couples to the two fermions via the Yukawa term
  $\bar{L} \, \phi \, N_R$ and in general will transform as
  $(\textbf{n}, Y)$ under $\rm SU(2)_L \otimes U(1)_Y$, setting $Y' = Y - 1/2$ and $\textbf{n}' = \textbf{n} \pm 1$. Of course,
  $\textbf{n}$ and $Y$ have to be correlated in such a way that an
  electrically neutral component of $\phi$ exists. For example, for
  $\textbf{n}=1$, only $Y=0$ is possible. For $\textbf{n}=2$, $Y= \pm
  1/2$, this being the case in which $\phi$ can be identified with a
  SM-like Higgs i.e. either $H$ or $H^c$. For $\textbf{n}=3$, $Y=\pm
  1, 0$. For $\textbf{n}=4$, $Y= \pm 1/2, \pm 3/2$ and so on.
\item The scalar whose VEV leads to the $\mu$-term will be denoted as
  $\varphi$. It will transform under $\rm SU(2)_L$ as $\textbf{1, 3,
  5, \dots 2n+1}$ and will have a hypercharge of $1-2\,Y$. Like in the
  case of $\phi$, the relation between the $\rm SU(2)_L \otimes
  U(1)_Y$ charges of $\varphi$ have to be such that a neutral
  component exists in order to avoid electric charge violation.
\end{itemize}

\begin{table}[h]
\centering
\begin{tabular}{| c || c | c | c || c | c | c  | }
  \hline 
& Fields    &    $\rm SU(2)_L \otimes U(1)_Y$   
&\hspace{.05cm}  $\rm U(1)_{B-L}$ \hspace{.05cm}$\to$\hspace{.05cm} $\mathbb{Z}_2$ \hspace{.05cm}  
& Fields    &    $\rm SU(2)_L \otimes U(1)_Y$   
&\hspace{.05cm}  $\rm U(1)_{B-L}$ \hspace{.05cm}$\to$\hspace{.05cm} $\mathbb{Z}_2$ \hspace{.05cm}     \\
\hline \hline
\multirow{4}{*}{ \begin{turn}{90} \hspace{0.9cm} \scriptsize{Fermions} \end{turn} }
&   $L_i$        	  &   ($\mathbf{2}, {-1/2}$)                    &  $ -1 \,\, \to \,\,  -1 $ 
&                     &                                             &               \\
&   $N_L$        	  &   ($\mathbf{n\pm 1}, {Y-1/2}$)              &  $ -1 \,\, \to \,\,  -1 $ 
&   $N_R$             &   ($\mathbf{n \pm 1}, {Y-1/2}$)             &  $ -1  \,\, \to \,\,  -1 $  \\
\hline \hline
\multirow{5}{*}{ \begin{turn}{90}\hspace{1.5cm} \scriptsize{Scalars} \end{turn} }
&   $H$  	         &  ($\mathbf{2}, {1/2})$                   	&  $ 0 \,\, \to \,\,  1 $ 
&   $\phi$           &  ($\mathbf{n}, Y)$                           &  $ 0 \,\, \to \,\,  1 $ \\
&   $\varphi$        &  ($\mathbf{1, 3, 5... 2n+1}, 1-2 \, Y)$     &  $ 2 \,\, \to \,\,  1 $     
&                    &                                              &               \\
    \hline
  \end{tabular}
\caption{\begin{footnotesize}Particle content of the generalized
    Majorana inverse seesaw. To generate a mass for the other SM
    fermions, a SM-like Higgs scalar $H$ is always needed, and we have
    explicitly included it in the list. In some particular cases,
    $\phi$ can be identified with the SM-like Higgs $H$ itself
    i.e. $\phi \equiv H$ and two separate particles are not
    required. The VEV of the scalar $\varphi$ breaks the $\rm B-L$
    symmetry and thus can be naturally small.
 \end{footnotesize}}
 \label{tab:su2majoranageneral}
\end{table}

Having established our notation, let us look at the possible multiplet
generalizations of the canonical inverse seesaw of
Section~\ref{sec:Majorana}. Since in this section we are restricting
ourselves to only multiplet generalizations, the particle content of
the models remain the same. Thus, in all cases we are led to the same
inverse seesaw formula in Eq.~\eqref{eq:MajoranaMass}. However, now
the fields can have more general $\rm SU(2)_L \otimes U(1)_Y$ charges,
as shown in Tab.~\ref{tab:su2majoranageneral}.

\begin{figure}[t!]
\centering
\includegraphics[scale=0.7]{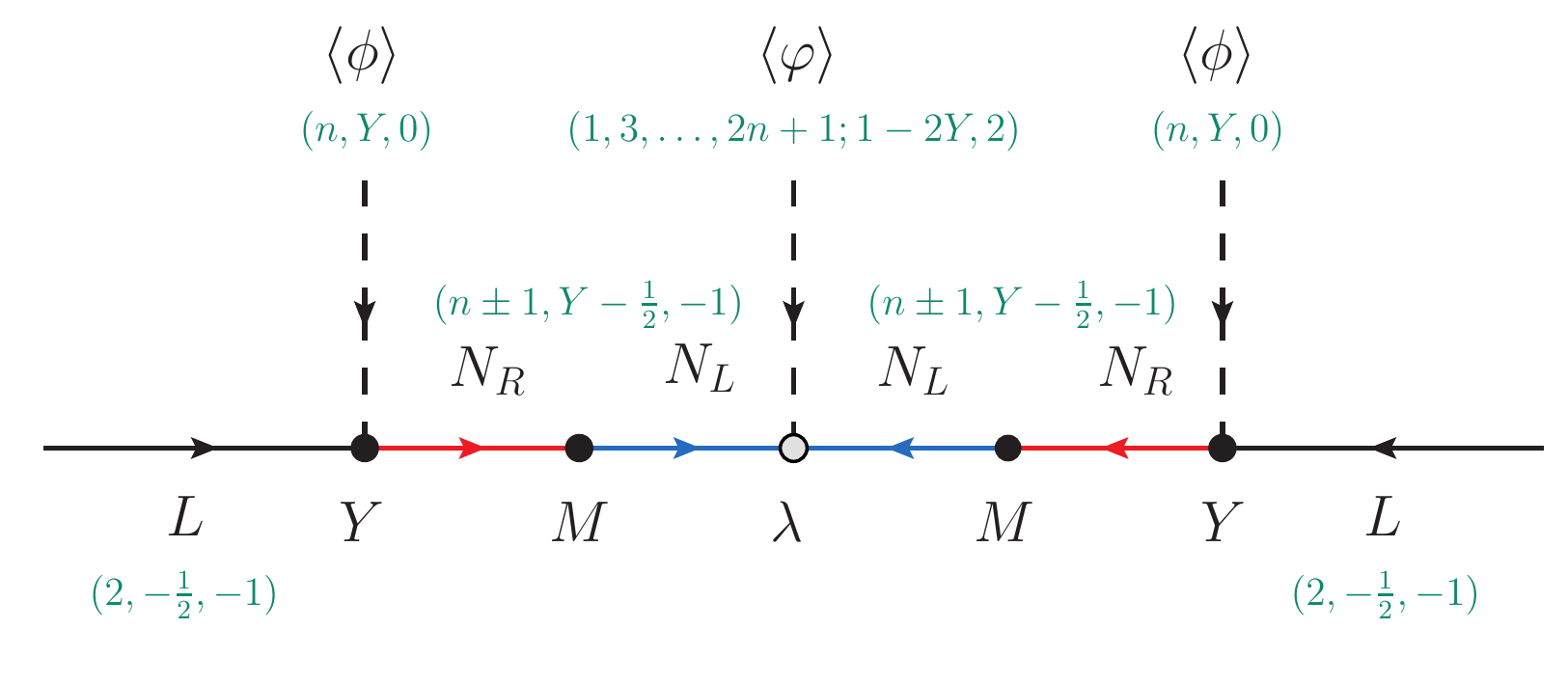}
\caption{Generalized Majorana inverse seesaw.
\label{fig:genmajo}}
\end{figure}

In Tab.~\ref{tab:su2majoranageneral}, apart from the scalars $\phi$
and $\varphi$, we have also explicitly added a SM Higgs-like $\rm SU(2)_L$
scalar doublet $H$. In models where $\phi$ has suitable quantum
numbers so as to be used for mass generation for quarks and charged
leptons, a separate $H$ field is not needed. In such cases one can
identify $\phi \equiv H$. The Feynman diagram for the inverse seesaw
generation of neutrino masses using general multiplets is shown in
Fig.~\ref{fig:genmajo}.

\begin{table}
\centering
\begin{tabular}{| c | c | c  | c | }
  \hline 
\hspace{1.5cm} Name of the model \hspace{1.5cm}   &    \hspace{1.5cm}  $\phi$     \hspace{1.5cm}  
&  \hspace{0.2cm}  $N_L$ and $N_R$  \hspace{0.2cm}                  & \hspace{0.9cm}  $\varphi$ ($\mu$-term)     \hspace{0.9cm}      \\
\hline \hline   
Type I inverse seesaw -  (2, 1, 0)           & $(\textbf{2}, 1/2) = H$   & (\textbf{1}, 0)    & (\textbf{1},0)   \\
\hline
\hline
Type III inverse seesaw - (2, 1, 0)          & $(\textbf{2}, 1/2) = H$   & (\textbf{3}, 0)    & (\textbf{1},0)   \\
\hline
Type III inverse seesaw - (2, 5, 0)          & $(\textbf{2}, 1/2) = H$   & (\textbf{3}, 0)    & (\textbf{5},0)   \\
\hline
Type III inverse seesaw - (2, 5, 2)         & (\textbf{2}, -1/2) = $H^c$ & (\textbf{3}, -1)    & (\textbf{5}, 2)   \\
\hline
Type III inverse seesaw - (4, 5, 1-2Y)       & (\textbf{4}, Y=-1/2, 3/2) & (\textbf{3}, Y-1/2)    & (\textbf{5}, 1-2Y)   \\
\hline
\hline
Type IV inverse seesaw - (3, 7, 1-2Y)        & (\textbf{3}, Y = 0, $\pm$ 1)      & $(\textbf{4}, Y-1/2)$    & $(\textbf{7}, 1-2Y)$   \\
\hline
\hline
Type V inverse seesaw - (4, 1, 0)            & (\textbf{4}, 1/2)  & (\textbf{5}, 0)    & (\textbf{1},0)   \\
\hline
Type V inverse seesaw - (4, 5 or 9, 0)       & (\textbf{4}, 1/2) & (\textbf{5}, 0)    & (\textbf{5} or \textbf{9}, 0)  \\
\hline
Type V inverse seesaw - (4, 5 or 9, 1-2Y)    & (\textbf{4}, Y=-1/2, 3/2) & (\textbf{5}, Y-1/2)    & (\textbf{5} or \textbf{9}, 1-2Y)   \\
\hline
Type V inverse seesaw - (4, 9, 4)           & (\textbf{4}, -3/2) & (\textbf{5}, -2)   & (\textbf{9}, 4)  \\
\hline
\hline
 \end{tabular}
 \caption{\begin{footnotesize} A few examples of the generalized
     inverse seesaw in the Majorana case. The nomenclature of the
     models is ``Type X inverse seesaw - (n, m, k)'' where X is the
     $\rm SU(2)_L$ multiplet of the fermion $N$, n and m are the $\rm
     SU(2)_L$ multiplets of $\phi$ and $\varphi$, respectively, and k
     is the hypercharge of $\varphi$. If the field $\phi$ is charged
     as $(\mathbf{2}, -1/2)$ under $\rm SU(2)_L \otimes U(1)_Y$ then
     it can be identified with the SM Higgs doublet $H$ while
     $\varphi$ is responsible for $\rm B-L$ breaking. We are not
     showing the models which generate a type-II seesaw
     contribution. See text for detailed
     discussion. \end{footnotesize}
 \label{tab:majoranaexamples}}
\end{table}

Some of the simplest generalized inverse seesaw models are listed in
Tab.~\ref{tab:majoranaexamples}. Some of the type-III models have been
discussed previously
in~\cite{Abada:2007ux,Gavela:2009cd,Ibanez:2009du,Ma:2009kh,Eboli:2011ia,Morisi:2012hu,Aguilar-Saavedra:2013twa,Law:2013gma}. In~\cite{Law:2013gma}
the first case of ``Type V inverse seesaw'' was also discussed. The
remaining cases, to the best of our knowledge, are being discussed for
first time by us. Let us now cover the simplest cases systematically.
 
\begin{enumerate}

 \item \textbf{For n = 1:} \\ The simplest case is obtained with
   $n=1$. In this case, the only option for the $Y$ hypercharge is $Y
   = 0$ as any other value will lead to electric charge breaking once
   $\phi$ gets a VEV.  Then the only option for the charges of $N_L,
   N_R$ is $(2,-1/2)$, which are the same as for the SM lepton doublet
   $L$. Finally, given these charges, $\varphi$ would be forced to
   transform as $(\textbf{3}, 1)$. However, one can see that the
   resulting model will also induce a type II seesaw which, depending
   on the parameter choices, can provide the leading order
   contribution to the light neutrino masses, among other
   phenomenological issues. Thus, this choice will at best lead to a
   mixed inverse-typeII seesaw and not a pure inverse seesaw as
   desired. Hence, we reject this possibility and will not discuss it
   further.
 \item \textbf{For n = 2:} \\ The next case is $n=2$, where we can
   identify $\phi$ with either $H$ (for $Y=1/2$) or $H^c$ (for
   $Y=-1/2$). Moreover, the internal fermions $N_R$ and $N_L$ can be
   either singlets or triplets under $\rm SU(2)_L$ and will carry a
   hypercharge equal to $Y-1/2$ i.e. either $0$ or $-1$. Several
   different possibilites arise here as we discuss now:
 
 \begin{enumerate}
 
  \item The case $Y=1/2$ with $N_L, N_R \sim (\textbf{1}, 0)$ implies
    $\varphi \sim (\textbf{1}, 0)$. It is nothing but the canonical
    Majorana inverse seesaw discussed in Sec.~\ref{sec:Majorana} and
    studied and extensively in the literature.
  \item The case $Y=-1/2$ with $N_L, N_R \sim (\textbf{1}, -1)$ does
    not lead to any viable neutrino mass model. This is because $N_L$
    and $N_R$ in this case have no electrically neutral components and
    hence are unsuitable to act as tree-level mediators for neutrino
    mass generation.
  \item Taking $N_L$ and $N_R$ to be two $\rm SU(2)_L$ triplets with
    hypercharge $0$, leads to three possibilities for $\varphi$:
    $\textbf{1, 3 or 5}$ under $\rm SU(2)_L$. Taking $\varphi \sim
    \textbf{1}$ under $\rm SU(2)_L$ leads to the ``type-III inverse
    seesaw'' shown in Tab.~\ref{tab:majoranaexamples}. The option
    $\varphi \sim \textbf{3}$ would lead to vanishing neutrino masses
    since two triplets contracting to another triplet is an
    antisymmetric combination, which vanishes if the two fields are
    identical. The last option $\varphi \sim \textbf{5}$ would be the
    ``type-III variant seesaw'' shown in
    Tab.~\ref{tab:majoranaexamples}.
  \item Finally, for $N_L, N_R \sim 3$ under $\rm SU(2)_L$ with $-1$
    hypercharge, the scalar $\varphi$ can only transform as
    $(\textbf{5}, 2)$. This is a novel possibility and would
    represent an exotic variant of the type-III inverse seesaw. In
    Tab.~\ref{tab:majoranaexamples} we refer to this possibility as
    ``exotic variant I'' of the type-III inverse seesaw.  As opposed
    to the 'normal' type-III inverse seesaw, this model will feature
    doubly electrically charged fermions and quadruply charged
    scalars.
 \end{enumerate}
 \item \textbf{For n = 3 :} In the case $n=3$ we have three
   possibilities for $Y$: $0$ and $\pm 1$. For each value of $Y$, the
   fermions can transform as either doublets or quadruplets under $\rm
   SU(2)_L$, with their hypercharge values ranging from $-3/2$ to
   $1/2$. There are several possible cases falling under this
   category. These are :
 \begin{enumerate}
  \item For $Y=1$, the fermions $N_L$ and $N_R$ will have a
    hypercharge of $1/2$. If they are $\rm SU(2)_L$ doublets, then
    $\varphi$ will again be an $\rm SU(2)_L$ triplet with hypercharge
    $1$. Thus, in this case, in addition to the inverse seesaw there
    will also be a type-II seesaw-like contribution. We therefore
    neglect this option. If the new fermions are quadruplets under
    $\rm SU(2)_L$, $\varphi$ will have hypercharge $-1$ and will
    transform as $\textbf{3}$ or $\textbf{7}$ under $\rm
    SU(2)_L$. However, the case $\varphi \sim (\textbf{3}, -1)$ would
    generate a type-II seesaw contribution. We call the only remaining
    posibility ``Type IV inverse seesaw'' in
    Tab.~\ref{tab:majoranaexamples}. Note that again due to the
    antisymmetry of the contractions $4 \times 4 \rightarrow 1$ and $4
    \times 4 \rightarrow 5$, the options of $\varphi$ transforming as
    $\textbf{1}$ and $\textbf{5}$ under $\rm SU(2)_L$ are both
    forbidden.
  \item For $Y=0$, again if the fermions are doublets then we will end
    up having a type-II seesaw contribution along with the inverse
    seesaw and as before we reject this possibility.
    If $N_L$ and $N_R$ are quadruplets of $\rm SU(2)_L$ with hypercharge $-1/2$
    then again $\varphi$ will transform as $\textbf{3}$ or
    $\textbf{7}$ with hypercharge of $1$. However, as before the case
    $\varphi \sim (\textbf{3}, 1)$ would generate a type-II seesaw
    contribution. So again the only viable possibility is $\varphi
    \sim \textbf{7}$ under $\rm SU(2)_L$ and the resulting model is again
    the ``Type IV inverse seesaw'' given in
    Tab.~\ref{tab:majoranaexamples}.
  \item For $Y=-1$ the new fermions will have a hypercharge of
    $-3/2$. Thus, they cannot be doublets as in that case they will
    not have any electrically neutral component to act as intermediate
    particles for neutrino mass generation. If the fermions are
    quadruplets, then the only option for $\varphi$ is to transform as
    $\textbf{7}$ under $\rm SU(2)_L$. In this case the possibility of it
    being a triplet will imply that $\varphi$ has no electrically
    neutral component and hence if it gets VEV, it would violate
    electric charge invariance.
  \item In summary, in all cases for $n = 3$, we only have one viable
    possibility, namely that of the fermions transforming as
    quadruplets of $\rm SU(2)_L$ with hypercharge $Y-1/2$, while
    $\varphi$ will transform as $(\textbf{7}, 1-2Y)$.
 \end{enumerate}
 
\item \textbf{For $n = 4$:}\\
   There are 4 possibilities for $Y$: $\pm
   1/2$ and $\pm 3/2$. The internal fermions will be either triplets
   or quintuplets, while $\varphi$ will be either $\textbf{1}$ (only
   for $Y=1/2$), $\textbf{5}$ or $\textbf{9}$ (only for the quintuplet
   case) under $\rm SU(2)_L$.
 
\begin{enumerate}

 \item The case in which $Y=1/2$ and the fermions are triplets would
   generate another contribution for the type-III inverse seesaw. We
   therefore neglect this possibility. If the fermions are quintuplets
   with hypercharge $0$, $\varphi$ could be either $(\textbf{1}, 0)$,
   leading to the ``type-V inverse seesaw'', or a variant of it, which
   we call ``type V variant'', with $\varphi$ transforming as either
   $(\textbf{5}, 0)$ or $(\textbf{9}, 0)$ as shown in
   Tab.~\ref{tab:majoranaexamples}.
 
 \item For $Y=-1/2$, the new fermions can be either quintuplets or
   triplets with hypercharge $-1$. $\varphi$ will transform as either
   a quintuplet, in both cases, or in the quintuplet case, as
   $\textbf{9}$ with hypercharge $2$. In
   Tab.~\ref{tab:majoranaexamples} these possibilites are listed as
   ``type V exoctic variant I''.
 
 \item If $Y = 3/2$ then the new fermions can be either triplets or
   quintuplets with hypercharge $1$. Again, $\varphi$ will transform
   as either a quintuplet, in both cases, or a $\textbf{9}$ only in
   the quintuplet fermion case, of hypercharge $-2$. This possibility
   is also listed as ``type V exoctic variant I'' in
   Tab.~\ref{tab:majoranaexamples}.
 
 \item Finally, the case $Y=-3/2$ only allows for quintuplet fermions
   of hypercharge $-2$. The only viable option for $\varphi$ is to
   transform as $(\textbf{9}, 4)$, leading to the ``type V exoctic
   variant II'' of Tab.~\ref{tab:majoranaexamples}.
 
\end{enumerate}

\item \textbf{For $n > 4$} : \\
  Higher values of $n$ are also
  possible. However, one can trivially generalize further to higher
  $n$ values and we will not discuss them explicitly.

\end{enumerate}

Before ending this section, we would like to comment on the advantages
of using spontaneous symmetry breaking rather than explicit breaking,
when exploring the model space for any given neutrino mass generation
mechanism. If one opts for explict symmetry breaking to generate the
$\mu$-term, one will miss many interesting models. This is because in
such case one can only consider $\mu$-terms that do not break gauge
symmetries, as gauge symmetries cannot be broken explicitly. This is
equivalent to restricting to only $\rm SU(2)_L$ singlet cases with
$Y=0$ in our analysis.
On the contrary, when the $\mu$-term is induced through spontaneous
symmetry breaking, the field $\varphi$ whose VEV will lead to the
$\mu$-term can in principle transform as any representation of the gauge
groups. In cases when $\varphi$ has non-trivial $\rm SU(2)_L \otimes
U(1)_Y$ transformations, its VEV will break gauge symmetry as
well. However, since it is spontaneous breaking and given the hierarchy
$\langle \varphi \rangle \ll v $; $v$ being the electroweak VEV, there
is no issue in such a breaking. Thus, generating the $\mu$-term
dynamically via spontaneous symmetry breaking reveals the full
landscape of models falling under the inverse seesaw mechanism.

\subsection{Generalized Dirac inverse seesaw}

\begin{figure}[t!]
\centering
\includegraphics[scale=0.7]{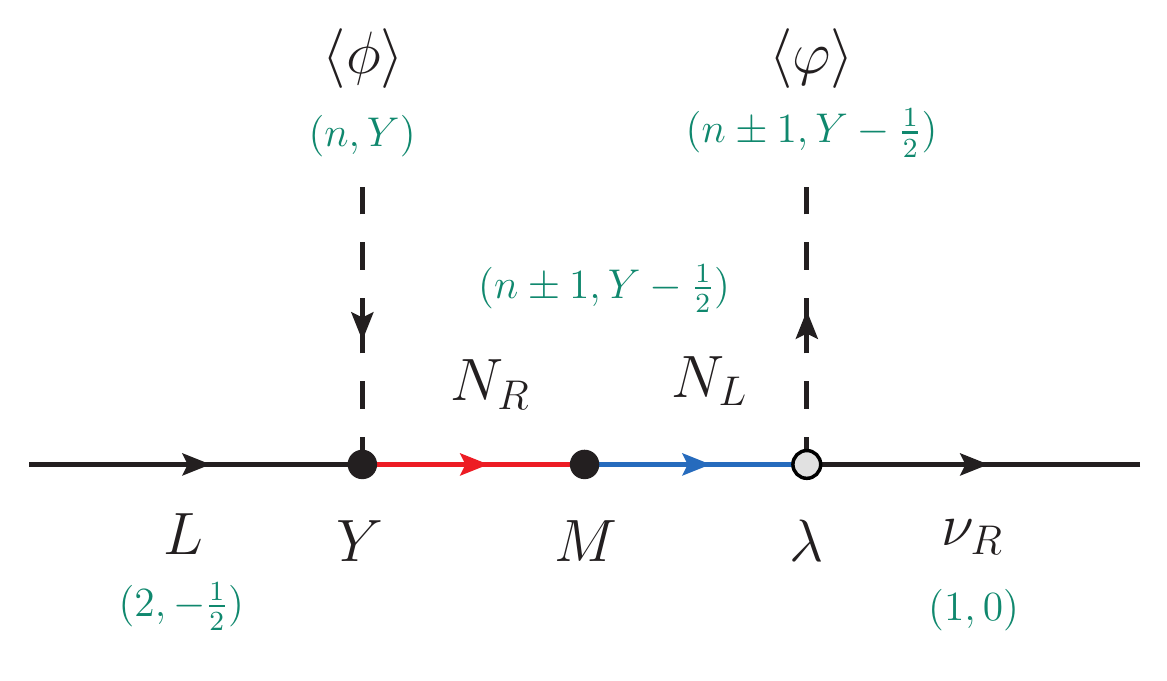}
\caption{Generalized Dirac inverse seesaw.
\label{fig:gendiracmin}}
\end{figure}

We now move on to the multiplet generalization of the Dirac inverse
seesaw. Again, the Lagrangian of the model would be formally identical
to the one in Sec.~\ref{subsec:dirac-inv}, but with higher $\rm
SU(2)_L$ multiplets and replacing $H \rightarrow \phi$ and $\chi \rightarrow \varphi$. The charges under $\rm U(1)_{B-L}$ will also be
identical and therefore the model would share the same appealing
features. The general $\rm SU(2)_L \times U(1)_Y$ charges can be seen
in Tab.~\ref{tab:generaldiracmin} and diagramatically in
Fig.~\ref{fig:gendiracmin}.

\begin{table}[h!]
\centering
\begin{tabular}{| c || c | c | c  || c | c | c |}
  \hline 
&   Fields            &  $\rm SU(2)_L \otimes U(1)_Y$           
& \hspace{0.05cm} $\rm U(1)_{B-L}$ \hspace{0.05cm} $\to$ \hspace{0.05cm} $\mathbb{Z}_3$ \hspace{0.05cm} 
&   Fields            &  $\rm SU(2)_L \otimes U(1)_Y$   
& \hspace{0.05cm} $\rm U(1)_{B-L}$ \hspace{0.05cm} $\to$ \hspace{0.05cm} $\mathbb{Z}_3$ \hspace{0.05cm}  \\                             
\hline \hline
\multirow{4}{*}{ \begin{turn}{90} \hspace{0.85cm} \tiny{Fermions} \end{turn} }
&   $L_i$        	  &   ($\mathbf{2}, {-1/2}$)       &    $-1 \,\, \to \,\, \omega^2$   	  
&   $\nu_R$        	  &   ($\mathbf{1}, {0}$)          &    $(-4, -4, 5) \, \to \, \omega^2$ \\
&   $N_L$             &   ($\mathbf{n \pm 1}, Y-1/2$)   &   $-1 \,\, \to \,\, \omega^2$     
&   $N_R$             &   ($\mathbf{n \pm  1}, Y-1/2$)  &   $-1 \,\, \to \,\, \omega^2$       \\
\hline \hline
\multirow{5}{*}{ \begin{turn}{90}\hspace{1.6cm} \tiny{Scalars} \end{turn} }
&   $H$  	         &  ($\mathbf{2}, {1/2}$)            &    $0 \,\, \to \,\, \omega^0 $      
&   $\varphi$        &  ($\mathbf{n \pm 1}, Y-1/2$)      &    $3 \,\, \to \,\, \omega^0 $        \\
&   $\phi$           &  ($\mathbf{n}, Y$)                &    $0 \,\, \to \,\, \omega^0 $     
& & & \\
     \hline
  \end{tabular}
\caption{\begin{footnotesize}Particle content of the generalized Dirac inverse
  seesaw. As in the Majorana case, here also $\phi$ can be identified with the SM Higgs
  under certain conditions. Again, $\varphi$ breaks the symmetry and
  generates the $\mu$-term. The particle charges under the residual $\mathbb{Z}_3$ symmetry are given by cube roots of unity with $\omega = e^{2\pi I/3}; \, \omega^3 = 1$.                                                                                                                                                                      \end{footnotesize}
\label{tab:generaldiracmin}}
\end{table}

The notation and conventions here are the same as in the previous Section~\ref{sec:generalizedmajorana}, which generalized the inverse seesaw mechanism in the Majorana case. Here we are again using the anomaly free $(-4,-4,5)$ solution for the $\rm U(1)_{B-L}$ symmetry with the $\rm B-L$ charges of $N_L, N_R$ assigned in a ``vector'' fashion such that the $\rm U(1)_{B-L}$ group remains anomaly free.
Note that for neutrinos to remain Dirac particles, an unbroken symmetry is needed to protect their Diracness. As in Section~\ref{sec:Dirac}, this role is again fullfiled by the unbroken residual $\mathbb{Z}_3$ subgroup of the $\rm U(1)_{B-L}$ symmetry, see Tab.~\ref{tab:generaldiracmin}.

Finally, let us list a few viable examples in
Tab.~\ref{tab:diracminexamples}. The general strategy to build these
models is similar to that discussed at length for the Majorana case in
Section~\ref{sec:generalizedmajorana}. We first fix the $\rm SU(2)_L
\otimes U(1)_Y$ charge of the scalar $\phi$. Depending on the $\phi$
charges, the $N_L$ and $N_R$ fields can have only one option for their
$\rm SU(2)_L \otimes U(1)_Y$ transformations. Finally, given the
charges of both $\phi$ and the $N_L,N_R$ fields, the viable options
for the $\varphi$ gauge charges can be obtained. The final list of
viable models for the Dirac inverse seesaw with generalized multiplets
are listed in Tab.~\ref{tab:diracminexamples}.

\begin{table}[h!]
\centering
\begin{tabular}{| c | c | c  | c | }
  \hline 
\hspace{0.2cm} Name of the model \hspace{0.2cm}   &  \hspace{0.2cm}  $N_L$ and $N_R$  \hspace{0.2cm}            &    \hspace{0.9cm}  $\phi$     \hspace{0.9cm}        & \hspace{0.9cm}  $\varphi$ ($\mu$-term)     \hspace{0.9cm}    \\
\hline \hline
Standard Dirac Inverse seesaw            &     $(\mathbf{1}, 0)$     &   $(\mathbf{2}, 1/2)$  &   $(\mathbf{1}, 0)$ \\
\hline \hline
Type-III Dirac Inverse seesaw            &     $(\mathbf{3}, 0)$     &   $(\mathbf{2}, 1/2)$  &   $(\mathbf{3}, 0)$ \\
\hline
Type-III Dirac Inverse seesaw variant I  &     $(\mathbf{3}, -1)$    &   $(\mathbf{2}, -1/2)$ &   $(\mathbf{3}, -1)$ \\
\hline 
Type-III Dirac Inverse seesaw variant II &     $(\mathbf{3}, 1)$     &   $(\mathbf{4}, 3/2)$  &   $(\mathbf{3}, 1)$ \\
\hline \hline
Exotic or Type IV Dirac inverse seesaw   &     $(\mathbf{4}, Y-1/2)$ &   $(\mathbf{3}, Y = 0, \pm 1)$ &   $(\mathbf{4}, Y-1/2)$ \\
\hline \hline
Type-V Dirac Inverse seesaw              &     $(\mathbf{5}, 0)$     &   $(\mathbf{4}, 1/2)$  &    $(\mathbf{5}, 0)$  \\
\hline
Type-V Dirac Inverse seesaw variant I    &     $(\mathbf{5}, 1)$     &   $(\mathbf{4}, 3/2)$  &   $(\mathbf{5}, 1)$  \\
\hline
Type-V Dirac Inverse seesaw variant II   &     $(\mathbf{5}, -1)$    &   $(\mathbf{4}, -1/2)$ &   $(\mathbf{5}, -1)$ \\
\hline
Type-V Dirac Inverse seesaw variant III  &     $(\mathbf{5}, -2)$    &   $(\mathbf{4}, -3/2)$ &   $(\mathbf{5}, -2)$ \\
\hline 
\hline

 \end{tabular}
 \caption{ A few examples of the generalized Dirac inverse seesaw.
 \label{tab:diracminexamples}}  
\end{table}

In Tab.~\ref{tab:diracminexamples} we have restricted outselves to $n=4$ i.e. upto the case in which $\phi$ transforms as a quadruplet under $\rm SU(2)_L$. Nevertheless, the generalization to higher $n > 4$ is rather straightforward.
Finally, as in the Majorana case, here also we have named the ``type'' of the model based on the $\rm SU(2)_L$ transformation of the $N_L, N_R$ fermions. The subclass numbering is based on the $\rm SU(2)_L$ transformation of $\phi$ as well as the $\rm SU(2)_L \otimes U(1)_Y$ transformation of the $\varphi$ field.

\section{Generalizing the inverse seesaw - II: Double inverse seesaw and beyond}
\label{sec:generalized2}

We will now move into a new type of generalization of the inverse
seesaw framework in which the fermionic sector of the model is
extended in order to obtain double or multiple $\mu$-term supression
of the neutrino mass.  As before, we start with the Majorana case,
showing how one can obtain a ``double inverse seesaw''. We then
discuss how one can generalize to triple and then multiple inverse
seesaw. After that we show that the same can be done for Dirac
neutrinos, explicitly working out the double and triple inverse
seesaws and ending with a discussion on the Dirac multiple inverse
seesaw.

\subsection{Majorana double inverse seesaw}

We start the discussion with the double inverse seesaw model for Majorana neutrinos. To build a double seesaw model, consider the canonical inverse seesaw model of Sec.~\ref{sec:Majorana} and add a new fermion $S_R$ with $\rm B-L$ charge $0$.
Moreover, let us change the charge of $\chi$ from $2$ to $1$. We keep
rest of the fields and their charges identical to those in the
canonical model. Therefore, the particle content and the charges of
the relevant fields are given in Tab.~\ref{tab:doublemajorana}.

\begin{table}[h!]
\begin{center}
\begin{tabular}{| c | c | c  || c | c | c |}
  \hline 
   Fields            &    $\rm SU(2)_{L} \otimes U(1)_Y$           
& \hspace{.05cm} $\rm U(1)_{B-L}$ \hspace{.05cm}
&   Fields            &   $\rm SU(2)_L \otimes U(1)_Y$            
& \hspace{.05cm} $\rm U(1)_{B-L}$ \hspace{.05cm}
\\
\hline \hline
   $L$               &   ($\mathbf{2}, {-1/2}$)        &    $-1 $      &   
   $S_R$             &   ($\mathbf{1}, {0}$)           &    $0 $        \\	
   $N_{L}$           &   ($\mathbf{1}, {0}$)           &    $-1$       &
   $N_{R}$           &   ($\mathbf{1}, {0}$)           &    $-1$        \\
\hline \hline
   $H$  	         &  ($\mathbf{2}, {1/2})$          &    $0$       &
   $\chi$ 	         &  ($\mathbf{1}, 0)$              &    $1$      \\
    \hline
  \end{tabular}
\end{center}
\caption{Particle content of the Majorana double inverse seesaw model. In this case, the $\rm U(1)_{B-L}$ symmetry gets completely broken after $\chi$ gets a VEV.
\label{tab:doublemajorana}}
\end{table}

\begin{figure}[t!]
\centering
\includegraphics[scale=0.7]{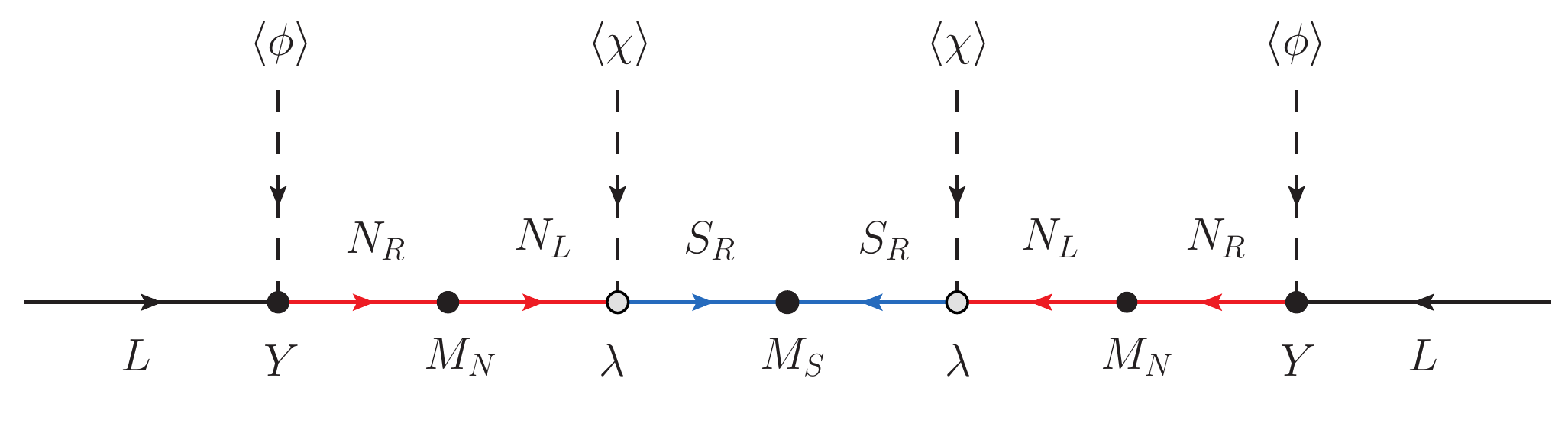}
\caption{Neutrino mass generation in the double Majorana inverse seesaw.
\label{fig:doublemaj}}
\end{figure}

The \SM \, and $\rm U(1)_{B-L}$ invariant Lagrangian relevant for
neutrino mass generation is given by
\begin{eqnarray}
 \mathcal{L}_{yuk} & = & Y \, \bar{L} H N_R \, + \, M_N \, \bar{N}_L N_R  \, + \, M_S \, \bar{S}_R S^c_R \, + \, \lambda \, \bar{N}_L \chi^* S_R\, + \, \lambda^\prime \, \bar{N}_R \chi S^c_R  \, + \, \rm{h.c.}
 \label{eq:lag-double-seesaw}
\end{eqnarray}
where $Y$, $\lambda$ and $\lambda^\prime$ are Yukawa couplings and $M_N$ and $M_S$ are gauge invariant mass terms. As shown diagrammatically in Fig.~\ref{fig:doublemaj}, neutrino masses are generated once the scalars get VEVs, $\langle H \rangle = v$ and $\langle \chi \rangle = u$.

The neutral fermion mass Lagrangian after symmetry breaking is given
in matrix form by
\begin{equation}
 \mathcal{L}_{m} \, = \, \left ( \begin{matrix}
        \bar{L}^c & \bar{N}_R & \bar{N}_L^c & \bar{S}_R
         \end{matrix}
 \right )
 \left ( \begin{matrix}
          0          & Y \, v      &    0      & 0  \\
          Y^T \,v    & 0           &    M_N    & \mu'   \\
          0          & M_N^T       &    0      & \mu  \\
          0          & \mu^{' T}    &   \mu^T   & M_S \\
          \end{matrix}
 \right )
 \left ( \begin{matrix}
          L \\
          N_R^c \\
          N_L \\
          S_R^c
         \end{matrix}
 \right ) \, ,
\end{equation}
where we have defined $\mu = \lambda u$ and $\mu' = \lambda^\prime u$. Curious readers will find this mass matrix a bit amusing as it does not look like an inverse seesaw mass matrix. However,  
this is simply because we have changed the ordering of the fields while writing the mass matrix. This is done so as to follow the sequence in which they appear in the fermionic line of Fig.~\ref{fig:doublemaj}.
Since the model has symmetry breaking $\mu$-terms as well as invariant terms, the model parameters naturally follow the inverse seesaw hierarchy
\begin{equation} \label{eq:seesaw2}
\mu, \mu' \ll Y \, v \ll M_N, M_S \, ,
\end{equation}
and the light neutrino mass matrix can be obtained in seesaw
approximation as
\begin{eqnarray}
 m_\nu &=&  \left ( \begin{matrix}
          Y v & 0 & 0
         \end{matrix}
 \right )
 \left ( \begin{matrix}

          0        &   M_N    &   \mu'   \\
          M_N^T    &   0      &   \mu  \\          
         \mu^{T '}   &   \mu^T    &   M_S \\
          \end{matrix}
 \right )^{-1}
 \left ( \begin{matrix}
          Y^T v \\
          0 \\
          0
         \end{matrix}
 \right ) \, .
\end{eqnarray}
Assuming one generation for the time being, this leads to the light
neutrino mass formula
\begin{equation}
 m_\nu = Y^2 \, v^2 \, \frac{\mu^2}{2 M_N \mu \mu'-M_N^2 \, M_S} \, \approx Y^2 \, v^2 \, \frac{\mu^2}{M_N^2 \, M_S}.
\end{equation}
One can easily see that this formula for light neutrino masses follows
the same spirit as the well-known formula in the canonical Majorana
inverse seesaw. However, now neutrino masses get suppressed by $\mu^2$
instead of $\mu$. This is the reason why we call this model \textit{double} inverse seesaw.

\subsection{Majorana triple inverse seesaw and beyond}

To go further along this idea, let us now add a new Weyl fermion and
rearrange the $\rm B-L$ charges. We will also need to add a new
symmetry-breaking scalar, see the relevant fields and charges in
Tab.~\ref{tab:triplemajorana}.

\begin{table}[h!]
\begin{center}
\begin{tabular}{| c | c | c  || c | c | c |}
  \hline 
   Fields            &      $\rm SU(2)_{L} \otimes U(1)_Y$     
&\hspace{.05cm}$\rm U(1)_{B-L}$ \hspace{.05cm}$\to$\hspace{.05cm} $\mathbb{Z}_2$\hspace{.05cm}      &   Fields            &     $\rm SU(2)_L \otimes U(1)_Y$            
&\hspace{.05cm}$\rm U(1)_{B-L}$ \hspace{.05cm}$\to$\hspace{.05cm} $\mathbb{Z}_2$\hspace{.05cm}  \\
\hline \hline
   $L$               &   ($\mathbf{2}, {-1/2}$)        &    $-1  \,\, \to \,\, -1$   	&   
                     &                                 &          \\	
   $N_{L}$           &   ($\mathbf{1}, {0}$)           &    $-1  \,\, \to \,\, -1$       &
   $N_{R}$           &   ($\mathbf{1}, {0}$)           &    $-1  \,\, \to \,\, -1$        \\
   $S_{L}$           &   ($\mathbf{1}, {0}$)           &  $a \,\, \to \,\, -1$       &
   $S_{R}$           &   ($\mathbf{1}, {0}$)           &  $a \,\, \to \,\, -1$        \\
\hline \hline
   $H$  	        &  ($\mathbf{2}, {1/2})$          &    $0 \,\, \to \,\, 1$       &
                    &                                 &          \\	
   $\chi_1$ 	    &  ($\mathbf{1}, 0)$              &   $a+1 \,\, \to \,\, 1$      &
   $\chi_2$ 	    &  ($\mathbf{1}, 0)$              &   $2a \,\, \to \,\, 1$      \\
    \hline
  \end{tabular}
\end{center}
\caption{Particle content of the Majorana triple inverse seesaw
  model. The $\rm U(1)_{B-L}$ symmetry gets broken into the residual
  $\mathbb{Z}_2$ after $\chi_1$ and $\chi_2$ get non-zero VEVs.
\label{tab:triplemajorana}}
\end{table}

To induce a triple inverse seesaw we need two different scalars carrying $\rm U(1)_{B-L}$ charges as shown in Tab.~\ref{tab:triplemajorana}, with $a$ being an integer. It should be noted that $\chi_1$ has to be different from $\chi_2$ in order for the triple inverse seesaw to provide the leading order contribution. Moreover, for the same reason, the $\rm B-L$ charges of both $\chi_1$ and $\chi_2$ need to be different from $0$.
The simplest solution is thus $a=2$
leading to $\chi_1 \sim 3$ and $\chi_2 \sim 4$.
Note that $a=1$ will lead to a type-I seesaw like contribution as the leading contribution. Hence, $a \neq 1$ is required to have triple inverse seesaw as the leading contribution to neutrino masses.

\begin{figure}[t!]
\centering
\includegraphics[scale=0.6]{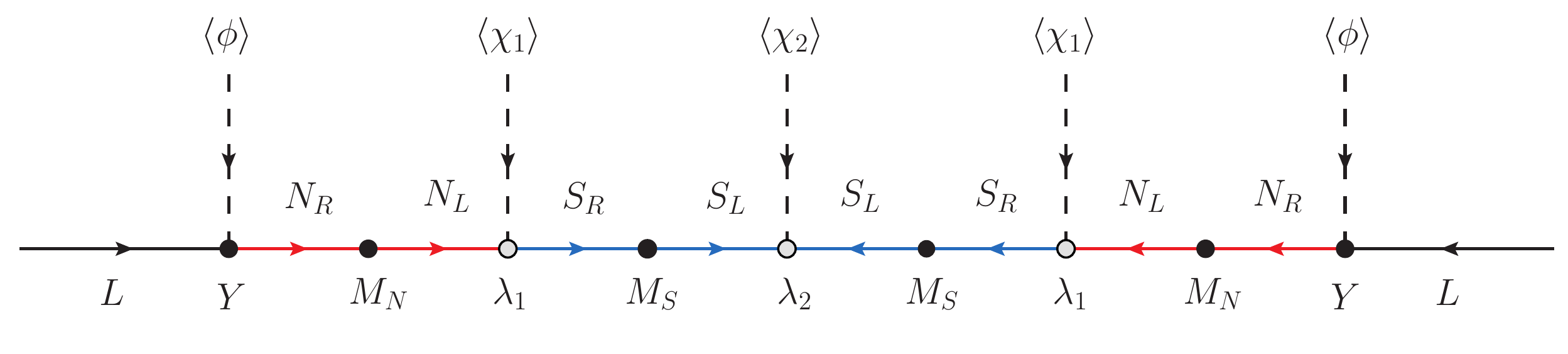}
\caption{Neutrino mass generation in the triple Majorana inverse
  seesaw. We note that the terms $N_R \, S_L$ and $S_R \, S_R$ cannot be
  avoided irrespective of the charge assignment. These terms generate
  $\mu'$ terms in analogy to the standard Majorana inverse seesaw.
\label{fig:triplemaj}}
\end{figure}

After symmetry breaking the light neutrinos become massive, as shown
in the diagram of Fig.~\ref{fig:triplemaj}. The resulting neutral
fermions mass Lagrangian in matrix form is given by
\begin{equation}
 \mathcal{L}_{m} \, = \, \left ( \begin{matrix}
        \bar{L}^c & \bar{N}_R & \bar{N}_L^c & \bar{S}_R & \bar{S}_L^c
         \end{matrix}
 \right )
 \left ( \begin{matrix}
          0          & Y \, v       &    0     & 0     &   0       \\
          Y^T \,v    & 0            &    M_N   & 0     &   \mu_1'   \\
          0          & M^T_N        &    0     & \mu_1 &   0         \\
          0          & 0            &   \mu^T_1& \mu_2'&   M_S        \\
          0          & \mu^{T '}_1  &    0     & M^T_S &   \mu_2       
          \end{matrix}
 \right )
 \left ( \begin{matrix}
          L \\
          N_R^c \\
          N_L \\
          S_R^c \\
          S_L
         \end{matrix}
 \right ) \, + \, h.c.
 \label{eq:3-mass-mat-maj}
\end{equation}
where $\mu_i = \lambda_i u_i$ and $\mu'_i = \lambda_i' u_i$; $i = 1,2$ with $\lambda_i, \lambda'_i$ being Yukawa couplings and $\langle \chi_i \rangle = u_i$ being the VEVs of the \SM \, singlet scalars. 
The model parameters naturally follow the already familiar inverse seesaw hierarchy
\begin{equation} \label{eq:seesaw3}
\mu_i, \mu_i' \ll Y \, v \ll M_N, M_S \, .
\end{equation}
Using the hierarchy of Eq.~\eqref{eq:seesaw3} in Eq.~\eqref{eq:3-mass-mat-maj} we obtain the light neutrino mass matrix in seesaw approximation as
\begin{eqnarray}
 m_\nu &=&  \left ( \begin{matrix}
          Y \, v & 0 & 0 & 0
         \end{matrix}
 \right )
 \left ( \begin{matrix}
         0       &    M_N   & 0     &   \mu_1'   \\
         M^T_N     &    0     & \mu_1 &   0         \\
         0       &    \mu^T_1 & \mu_2'&   M_S        \\
         \mu^{T ' }_1  &    0     & M^T_S   &   \mu_2       
          \end{matrix}
 \right )^{-1}
 \left ( \begin{matrix}
         Y^T \, v \\
          0 \\
          0 \\
          0
         \end{matrix}
 \right ) \, .
\end{eqnarray}
Assuming one generation, one obtains the light neutrino mass formula
\begin{equation}
 m_\nu = Y^2 v^2 \, \frac{\mu_1^2 \mu_2}{M_N^2 M_S^2 - 2 M_S M_N \mu_1 \mu_1'+ \mu_1^2 \mu_1'^2 - M_N^2 \mu_2 \mu_2'}  \approx  Y^2 v^2 \, \frac{\mu_1^2 \, \mu_2}{M_N^2 M_S^2} \, .
\end{equation}
In direct analogy with the standard inverse seesaw, contributions
coming from $\mu'$ can be safely neglected. Note that here the
suppression mechanism is enhanced by $\mu_1^2 \mu_2$. Therefore, we
would call this mechanism \textit{triple} Majorana inverse seesaw.

Further developments into quadruple, quintuple, \dots, Majorana
inverse seesaw mechanisms are straightforward.
One indeed has to ensure that the $n$ th order inverse seesaw is the leading order contribution to neutrino masses. This can always be ensured by choosing appropriate charges for the particles transforming under the symmetry whose breaking leads to the $\mu$-terms of the model. 
Having ensured that, in general we find 
\begin{itemize}
 \item \textbf{For $\boldsymbol{(2n-1)}$th order Inverse Seesaw:} \\
   In this case we require ``$n$ pseudo-vector pairs'' of fermions and ``$n$'' scalars. The resulting leading neutrino masses for a $(2n-1)$th order inverse seesaw (one generation) are given by
\begin{equation}
 m_\nu \approx Y^2 \, v^2 \, \frac{\mu_{n}}{M_n^2} \, \prod_{i=1}^{i = n-1} \, \frac{\mu_i^2}{M_i^2} \, ,
\end{equation}
where $Y$ is the Yukawa coupling involving the lepton doublet, $\mu_i$; $i = 1, \cdots,n$ are the small symmetry breaking $\mu$-terms and $M_i$; $i = 1, \cdots,n$ are the pseudo-Dirac masses for the pseudo-vector pairs of fermions.
\item \textbf{For $\boldsymbol{2n}$th order Inverse Seesaw:} \\
 In this case we require ``$n$ pseudo-vector fermion pairs'', a chiral fermion and ``$n$'' scalars. The resulting leading contributions to neutrino masses for a $2n$ th order inverse seesaw (one generation) are given by
 \begin{eqnarray}
 m_\nu & \approx &  Y^2 \, v^2 \, \frac{1}{M_{n+1}} \, \prod_{i=1}^{i = n} \, \frac{\mu_i^2}{M_i^2} \, ,
 \end{eqnarray}
where the definitions of all the couplings are the same as in the previous case.
\end{itemize}
%

\subsection{Dirac ``Double'' Inverse Seesaw}
\label{subsec:dirac-double}

To build the double Dirac inverse seesaw, we start with the field and
symmetry inventory of the minimal Dirac inverse seesaw of
Section~\ref{sec:Dirac}. To it, we add new VL fermions,
$S_{L,R}$, and a new singlet scalar $\chi_2$. The only modification in
the $\rm B-L$ charges is that we will take $(\chi_1, \chi_2)$ to
transform as $(\chi_1, \chi_2) \sim (6, -9)$. Remember that in the
previous example we had $\chi \sim 3$. Moreover, we take the new
VL fermion to transform as $S_{L, R} \sim 5$, while the rest
of the fields share their transformation properties with the previous
model as shown in Tab.~\ref{tab:diracmodel1}.

With the above choice of $\rm U(1)_{B-L}$ charges it is easy to check
that the model is anomaly free and can be gauged if desired. Also, the
$\rm B-L$ charges of the scalars are chosen in such a way that their
VEVs break $\rm U(1)_{B-L} \to \mathbb{Z}_3$. This residual
$\mathbb{Z}_3$ symmetry remains unbroken thus ensuring the Dirac
nature of neutrinos.
Note that the fields $\nu_R$ and $S_R$ share the same transformation
properties, but we call them differently to follow the notational
conventions of the previous and following sections.

\begin{table}[h!]
\centering
\begin{tabular}{| c || c | c | c || c | c |  c | }
  \hline 
& Fields   &   $\rm SU(2)_L \otimes U(1)_Y$   
&\hspace{.05cm}$\rm U(1)_{B-L}$\hspace{.05cm}$\to$\hspace{.05cm}$\mathbb{Z}_3$\hspace{.05cm}  
& Fields   &   $\rm SU(2)_L \otimes U(1)_Y$   
&\hspace{.05cm}$\rm U(1)_{B-L}$\hspace{.05cm}$\to$\hspace{.05cm}$\mathbb{Z}_3$\hspace{.05cm} 
\\
\hline \hline
\multirow{4}{*}{ \begin{turn}{90} \hspace{0.15cm} Fermions \end{turn} }
&   $L_i$        	  &   ($\mathbf{2}, {-1/2}$)       &    $-1 \,\, \to \,\, \omega^2$
&   $\nu_R$        	  &   ($\mathbf{1}, {0}$)          &$(-4, -4, 5) \,\, \to \,\, \omega^2$   \\	
&   $N_L$        	  &   ($\mathbf{1}, {0}$)          &    $-1 \,\, \to \,\, \omega^2$
&   $N_R$        	  &   ($\mathbf{1}, {0}$)          &    $-1 \,\, \to \,\, \omega^2$   \\	
&   $S_L$        	  &   ($\mathbf{1}, {0}$)          &    $5 \,\, \to \,\, \omega^2$
&   $S_R$        	  &   ($\mathbf{1}, {0}$)          &    $5 \,\, \to \,\, \omega^2$    \\	
\hline \hline
\multirow{5}{*}{ \begin{turn}{90}\hspace{1.55cm} Scalars \end{turn} }
&   $H$          	  &   ($\mathbf{2}, {1/2}$)       &    $0 \,\, \to \,\, \omega^0$
&   $\chi_1$          &   ($\mathbf{1}, {0}$)         &    $6 \,\, \to \,\, \omega^0$   \\   
&   $\chi_2$          &   ($\mathbf{1}, {0}$)         &    $-9 \,\, \to \,\, \omega^0$
&   &  &   \\	

    \hline
  \end{tabular}
\caption{Particle content of the Dirac analogue of the inverse
  seesaw. The $\rm U(1)_{B-L}$ charges of the fermions are fixed by an anomaly cancellation condition while the $\rm U(1)_{B-L}$ charges of the scalars are chosen such that the residual $\mathbb{Z}_3$ symmetry remains unbroken and the leading contribution to neutrino mass is induced by the double inverse seesaw.
\label{tab:diracmodel1}}
\end{table}

\begin{figure}[t!]
\centering
\includegraphics[scale=0.7]{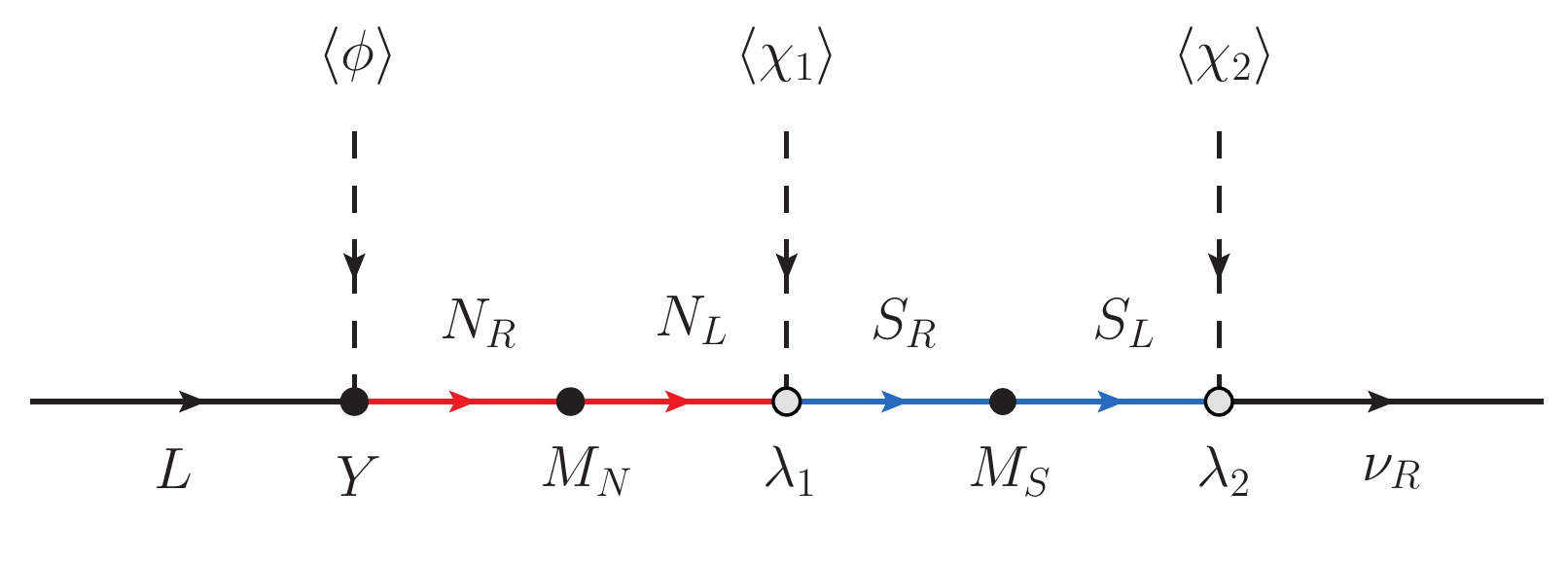}
\caption{Neutrino mass generation in the Dirac double inverse
  seesaw. The $\mu$-terms can be either explicit or spontaneously
  generated by the VEV of the scalars $\chi_1$ and $\chi_2$.
\label{fig:Dirac}}
\end{figure}

With all these ingredients, the Lagrangian of the model relevant to
neutrino mass generation is given by
\begin{align}
\label{eq:diracgenerallag}
\mathcal{L}_{\rm Dir} \, = & \, Y \, \bar{L} \, \widetilde H \, N_R \, + \, \lambda_2 \, \bar{S}_L \, \chi_2^* \, \nu_R \, + \, \lambda_1 \, \bar{N}_L \,  \chi_1^* \, S_R  + \lambda^\prime_1 \, \bar{S}_L \, \chi_1 \, N_R \, \\
& \, + \, M_N \, \bar{N}_L N_R \, + \, M_S \, \bar{S}_L S_R \, + \hc \, . 
\end{align}
Symmetry breaking is triggered by the scalar VEVs
\begin{equation}
\langle H \rangle = v \quad , \quad \langle \chi_1 \rangle = u_1 \quad , \quad \langle \chi_2 \rangle = u_2 \, ,
\end{equation}
which lead to the following $\mu$-terms,
\begin{equation}
\mu_1 = \lambda_1 \, u_1 \quad , \quad \mu^\prime_1 = \lambda^\prime_1 \, u_1 \, \quad , \quad \mu_2 = \lambda_2 \, u_2 \, .
\end{equation}
After symmetry breaking, Eq.~\eqref{eq:diracgenerallag} leads to the
mass Lagrangian in matrix form
\begin{equation}
 \mathcal{L}_{m} = \left ( \begin{matrix}
          \bar{\nu}_L & \bar{N}_L & \bar{S}_L
         \end{matrix}
 \right )
 \left ( \begin{matrix}
          0               & Y_N \, v     & 0 \\
          0               & M_N          & \mu_1 \\
          \mu_2      & \mu^\prime_1        & M_S          
          \end{matrix}
 \right )
 \left ( \begin{matrix}
          \nu_R \\
          N_R \\
          S_R
         \end{matrix}
 \right ) \, .
 \label{eq:double-dirac-lag}
\end{equation}
Again, the natural hierarchy among the parameters of the model is
\begin{equation}
\mu_1 , \mu^\prime_1, \mu_2 \ll Y \, v \ll M_N , M_S \, ,
\end{equation}
which leads to the light neutrino mass matrix
\begin{eqnarray}
 m_\nu &=&  \left ( \begin{matrix}
          Y \, v & 0
         \end{matrix}
 \right )
 \left ( \begin{matrix}

          M_N      & \mu_1\\
          \mu_1^\prime    & M_S
          \end{matrix}
 \right )^{-1}
 \left ( \begin{matrix}
          0 \\
          \mu_2
         \end{matrix}
 \right ) \, .
\end{eqnarray}
For one generation of light neutrinos, this is equivalent to
\begin{equation} \label{eq:2-DiracMass}
 m_\nu = Y_N \, \frac{v \, \mu_1 \, \mu_2}{\mu_1 \mu^\prime_1 - M_N M_S} \simeq - Y_N  \, v \, \frac{\mu_1 \mu_2}{M_N \, M_S} \, .
\end{equation}
This result is illustrated in the Feynman diagram shown in
Fig.~\ref{fig:Dirac}.

As it is clear from Eq.~\eqref{eq:2-DiracMass}, neutrino masses in
this case are suppressed by two $\mu$-terms and hence the name Dirac
\textit{double} inverse seesaw.
Note that the effect of $\mu^\prime$ is subleading in the neutrino mass
generation. 
Finally, we should remark that the spontaneous breaking of $\rm U(1)_{B-L}$ by the $u_1$ and $u_2$ VEVs leaves a residual $\mathbb{Z}_{3}$ symmetry.  As in the minimal Dirac model of Section~\ref{sec:Dirac}, here too all scalars transform trivially under this residual symmetry, while all fermions (except quarks which transform as $\omega$)  transform as
$\omega^2$. Again, this symmetry forbids all Majorana mass terms and
protects the Diracness of light neutrinos.

\subsection{Dirac triple inverse seesaw and beyond}

In order to build the triple Dirac inverse seesaw, we take the same
field inventory as in Sec.~\ref{subsec:dirac-double} and add a new set
of VL fermions, which we denote as $T$. Moreover, we need to
add a new scalar and modify the $\rm B-L$ symmetry charges of the
symmetry breaking scalars.
The $\rm B-L$ anomaly free 445-solution can still be implemented, with
the new fermions $S_{L,R}, T_{L,R}$ getting ``vector'' $\rm B-L$
charges, thus preserving the anomaly free structure, see
Tab.~\ref{tab:tripledirac}.

\begin{table}[h!]
\centering
\begin{tabular}{| c || c | c | c  || c | c | c | }
  \hline 
&   Fields            &     \hspace{0.2cm} $\rm SU(2)_L \otimes U(1)_Y$           
&\hspace{.05cm}$\rm U(1)_{B-L}$ \hspace{.05cm}$\to$\hspace{.05cm} $\mathbb{Z}_3$\hspace{.05cm}       &   Fields            &     \hspace{0.2cm} $\rm SU(2)_L \otimes U(1)_Y$           
&\hspace{.05cm} $\rm U(1)_{B-L}$ \hspace{.05cm}$\to$\hspace{.05cm} $\mathbb{Z}_3$\hspace{.05cm}                       \\
\hline \hline
\multirow{4}{*}{ \begin{turn}{90} \hspace{-0.4cm} Fermions \end{turn} }
&   $L_i$        	  &   ($\mathbf{2}, {-1/2}$)   &    $-1 \,\, \to \,\, \omega^2$   	    	 
&   $\nu_R$        	  &   ($\mathbf{1}, {0}$)      & $(-4, -4, 5) \,\, \to \,\, \omega^2$ \\	
&   $N_L$        	  &   ($\mathbf{1}, {0}$)      &    $-1 \,\, \to \,\, \omega^2$   	    	
&   $N_R$        	  &   ($\mathbf{1}, {0}$)      &    $-1 \,\, \to \,\, \omega^2$    	  \\	
&   $S_L$        	  &   ($\mathbf{1}, {0}$)      &    $-4-b \,\, \to \,\, \omega^2$  
&   $S_R$        	  &   ($\mathbf{1}, {0}$)      &    $-4-b \,\, \to \,\, \omega^2$    \\	
&   $T_L$        	  &   ($\mathbf{1}, {0}$)      &    $a-1 \,\, \to \,\, \omega^2$  	  	  
&   $T_R$        	  &   ($\mathbf{1}, {0}$)      &    $a-1 \,\, \to \,\, \omega^2$   \\	
\hline \hline
\multirow{5}{*}{ \begin{turn}{90}\hspace{1.55cm} Scalars \end{turn} }
&   $H$          	  &   ($\mathbf{2}, {1/2}$)       &    $0 \,\, \to \,\, \omega^0$   
&   $\chi_1$          &   ($\mathbf{1}, {0}$)         &    $a \,\, \to \,\, \omega^0$   \\	
&   $\chi_2$          &   ($\mathbf{1}, {0}$)         &    $b \,\, \to \,\, \omega^0$   
&   $\chi_3$          &   ($\mathbf{1}, {0}$)         &    $-3-a-b \,\, \to \,\, \omega^0$  \\	
    \hline
  \end{tabular} 
\caption{Particle content of the triple Dirac inverse seesaw. The free charges $a$ and $b$ can only take certain values, see text for a more detailed discussion.
  \label{tab:tripledirac}}
\end{table}

In this construction we must again emphasize the importance of the
correct symmetry breaking pattern and the residual unbroken
symmetry. If the parameters $a,b$ are chosen without care, one may
generate Majorana mass terms which would in turn spoil the Dirac
nature of neutrinos. For example, the choice $a=2$ would allow the
presence of the mass term $\bar{N}_L^c N_L \chi_2$, which would
eventually induce Majorana neutrino masses through the effective
operator $\bar{L}^c L H H \chi_2$.
As a general rule, the three scalars $\chi_i$; $i = 1,2,3$ under $\rm
U(1)_{B-L}$ must transform as multiples of $3$, the sum of their
charges must be $3$ and none of them should have $0$ charge under the
$\rm U(1)_{B-L}$ symmetry.

\begin{figure}[t!]
\centering
\includegraphics[scale=0.7]{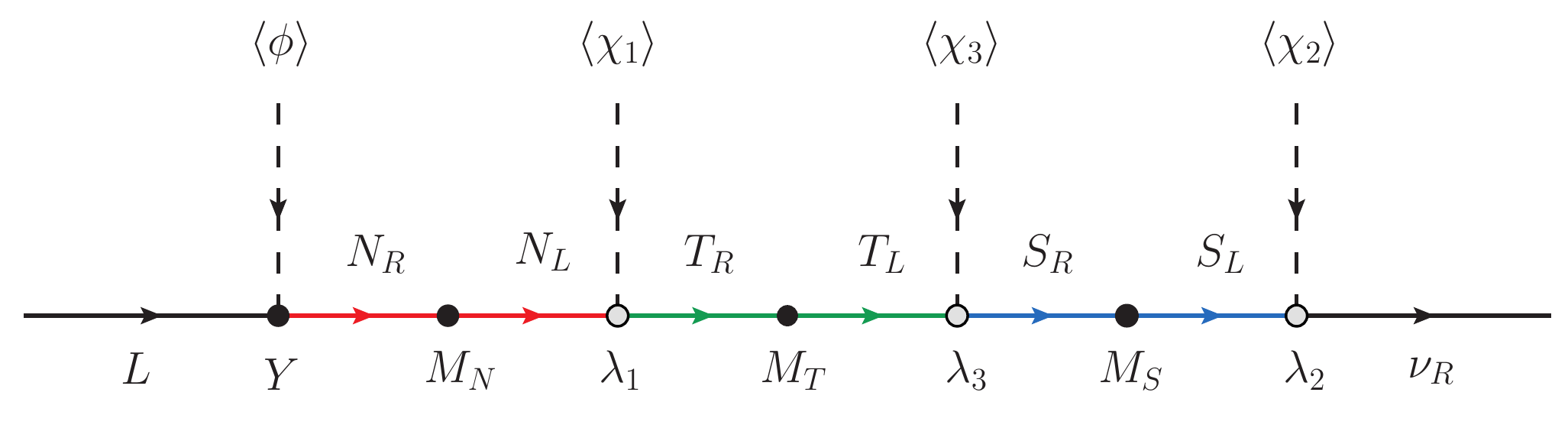}
\caption{Neutrino mass generation in the Dirac triple inverse
  seesaw. The terms $N_R T_L$ and $T_R S_L$ cannot be avoided
  irrespective of the charge choices. These terms generate $\mu'$
  terms in analogy to the standard Majorana inverse seesaw, but they
  play a subleading role in the generation of neutrino masses.
\label{fig:doubledir}}
\end{figure}

Neutrinos get massive after symmetry breaking, as shown in the diagram
of Fig.~\ref{fig:doubledir}. The neutral fermions mass Lagrangian is
given by
\begin{equation}
 \mathcal{L}_{m} = \left ( \begin{matrix}
          \bar{\nu}_L & \bar{N}_L & \bar{S}_L & \bar{T}_L
         \end{matrix}
 \right )
 \left ( \begin{matrix}
          0               & Y_N \, v     & 0    &   0    \\
          0               & M_N          & 0    &   \mu_2 \\
          \mu_1           & 0            & M_S  &   \mu_3' \\
          0               & \mu_2'       & \mu_3&   M_T
          \end{matrix}
 \right )
 \left ( \begin{matrix}
          \nu_R \\
          N_R \\
          S_R \\
          T_R
         \end{matrix}
 \right ) \, .
\end{equation}
Here all $\mu$-terms are defined as $\mu_i = Y_i u_i$ and $\mu_i' =
Y_i' u_i$. Again, the natural hierarchy among the parameters of the
model is
\begin{equation}
\mu_i , \mu_i^\prime \ll Y_N \, v  \ll M_N , M_S, M_T \, ,
\end{equation}
which leads to the light neutrino mass matrix
\begin{eqnarray}
 m_\nu &=&  \left ( \begin{matrix}
          Y_N \, v & 0 & 0
         \end{matrix}
 \right )
 \left ( \begin{matrix}

          M_N      & 0     & \mu_2 \\
          0        & M_S   & \mu_3' \\
          \mu_2'   & \mu_3 & M_T
          \end{matrix}
 \right )^{-1}
 \left ( \begin{matrix}
          0 \\
          \mu_1 \\          
          0          
         \end{matrix}
 \right ) \, .
\end{eqnarray}
For one generation of light neutrinos this is equivalent to
\begin{equation} \label{eq:DoubleDiracMass}
 m_\nu = Y_N \, v \, \frac{\mu_1 \mu_2 \mu_3}{M_N M_S M_T - M_S \mu_2 \mu_2' - M_N \mu_3 \mu_3'} \simeq Y_N \, v \, \frac{\mu_1 \mu_2 \mu_3}{M_N M_S M_T}
\end{equation}
Note that again the effect of $\mu_i^\prime$ is subleading in the
neutrino mass formula due to the assumed inverse seesaw hierarchy.

We finally point out that in order to induce a quadruple Dirac inverse
seesaw and beyond one would just need to sequentially add a new
VL fermion alongside a new scalar with a judicious choice of
symmetry breaking charges for the scalars. 
To obtain an $n$th order Dirac inverse seesaw as the leading contribution we need to add $n$ VL fermionic pairs and $n$ scalars. Of course appropriate symmetries, with particles carrying appropriate charges  under them, are required to ensure that neutrinos are Dirac particles with the leading contribution to their mass given by the $n$th order Dirac inverse seesaw as
\begin{eqnarray}
 m_\nu & \approx & Y \, v \, \prod_{i=1}^{i = n} \, \frac{\mu_i}{M_i} \, ,
\end{eqnarray}
where as before  $Y$ is the Yukawa coupling involving the lepton doublet, $\mu_i$; $i = 1, \cdots,n$ are the small symmetry breaking $\mu$-terms and $M_i$; $i = 1, \cdots,n$ are the masses for the vector pairs of fermions.
\\

A combination of the generalizations shown in
Sec.~\ref{sec:generalized1} and Sec.~\ref{sec:generalized2} is
straightforward and will not be developed here.

\section{Summary and conclusions}
\label{sec:summary}

To summarize, in this work we have developed the idea of inverse seesaw to encompass a whole class of mass generation mechanisms for both Majorana and Dirac neutrinos. We began by reviewing the famous canonical inverse seesaw before developing its Dirac analogue. We then showed that the idea of inverse seesaw is very general and can be implemented in a variety of ways. In particular we focused on two distinct type of extensions. In Section~\ref{sec:generalized1} we focused on developing the ``multiplet'' extensions of the inverse seesaw. We showed that both the canonical Majorana inverse seesaw and its Dirac analogue can be generalized by using fermions and scalars which transform as higher $\rm SU(2)_L$ multiplets with appropriate hypercharges. Subsequently in Section~\ref{sec:generalized2} we discussed the ``multiple $\mu$'' extensions.
We showed that one can generalize the idea of inverse seesaw to models where the neutrino mass is suppressed
by multiple symmetry breaking $\mu$-terms. We explicitly constructed doubly and triply suppressed inverse seesaw models for both Majorana and Dirac neutrinos. This idea can in fact be generalized in a rather straightforward manner to higher order inverse seesaw models. The general mass for neutrinos in an $n$th order inverse seesaw models along with the minimal set of new particles required in such models is summarized below in Tab.~\ref{tab:summary}. 
{
\renewcommand{\arraystretch}{1.6}
\begin{table}[h!]
\begin{center}
\begin{tabular}{| c | c | c | c  |}
  \hline 
  Model & $m_\nu$ formula            &    New fermions &  New scalars      \\
  \hline
  Majorana $(2n-1)$th inverse seesaw & $Y^2 v^2 \frac{\mu_{n}}{M_n^2}\prod_{i=1}^{i = n-1}\frac{\mu_i^2}{M_i^2}$ & $n$ VL pairs & $n$ scalars \\
  \hline
  Majorana $2n$th inverse seesaw & $Y^2 v^2 \frac{1}{M_{n+1}}\prod_{i=1}^{i = n}\frac{\mu_i^2}{M_i^2}$ & $n$ VL pairs and a Weyl fermion & $n$ scalars \\
  \hline
    Dirac $n$th inverse seesaw & $Y v \prod_{i=1}^{i = n}\frac{\mu_i}{M_i}$ & $n$ VL pairs & $n$ scalars \\
    \hline
  \end{tabular}
\end{center}
\caption{The neutrino mass formula and the minimal set of fields required for an $n$th order inverse seesaw model. Note that this table is written assuming one generation of neutrinos. The mass for all three generations can be generated simply requiring multiple generations for each type of the new fermions.}
\label{tab:summary}
\end{table}
}

It is noted that a combination of both techniques is also straightforward and hence not explicitly pursued in this work. Following the ideas above, it is possible to construct the 'multiple $\mu$' inverse seesaw models in a framework with higher order multiplets of $\rm SU(2)_L \times U(1)_Y$. 

Additionally let us emphasize the central role of the $\rm B-L$ symmetry both in the Majorana and the Dirac models. It is clear that these models need a symmetry argument in order to properly realize the inverse mechanism and, while not the only option, $\rm B-L$ represents a natural, minimal and elegant option which, in the Dirac case, can also be used to ensure the Diracness of neutrinos. Moreover, by using the anomaly free ``\textit{445-solution}'', we have ensured that all Dirac models are anomaly free and can therefore be gauged.

Finally  the models developed here are expected to have novel and interesting phenomenological signatures, both in colliders as well as in low scale experiments like those looking for signatures of lepton flavour or lepton number breaking. We plan to systematically explore these in follow up works.

\section*{Acknoledgements}

Work supported by the Spanish grants FPA2017-85216-P
(MINECO/AEI/FEDER, UE), SEJI/2018/033 (Generalitat Valenciana) and
FPA2017-90566-REDC (Red Consolider MultiDark), PROMETEO/2018/165 (Generalitat  Valenciana).  AV acknowledges financial support from MINECO through the Ramón y Cajal contract
RYC2018-025795-I. The work of RS is supported by the SERB, Government of India  grant under the file number SRG/2020/002303. The work of S.C.Ch.\ is supported by the Spanish FPI grant BES-2016-076643.


\bibliographystyle{utphys}
\bibliography{refs}

\end{document}